\documentclass{aa}  

\usepackage{makecell}
\usepackage{graphicx}
\usepackage{natbib}
\usepackage[utf8]{inputenc}
\usepackage{natbib}
\usepackage[mathscr]{euscript}
\usepackage{lscape}
\usepackage[dvipsnames]{xcolor}
\definecolor{myblue}{RGB}{63,200,244}
\definecolor{ocre}{RGB}{52,177,201}

\begin{document}

  \title{Neutron-capture elements in dwarf galaxies~II: \\Challenges for the $s$- and $i$-processes at low metallicity\thanks{Based on VLT/FLAMES observations collected at the European Organisation for Astronomical Research (ESO) in the Southern Hemisphere under programmes 71.B-0641, 171.B-0588, and 092.B-0194(A).}}
\titlerunning{Challenges for the $s$- and $i$-processes at low metallicity}
   
   \author{\'{A}.~Sk\'{u}lad\'{o}ttir
   		\inst{1,2,3}
        \and
        C.~J.~Hansen\inst{1}
       	\and
       	A.~Choplin\inst{4}
       	\and
       	S.~Salvadori \inst{2,3}
       	\and
       	M.~Hampel \inst{5}
       	\and
       	S.~W.~Campbell \inst{5}
       	}
          
   \institute{
   			Max-Planck-Institut f$\ddot{\text{u}}$r Astronomie, K$\ddot{\text{o}}$nigstuhl 17, D-69117 Heidelberg, Germany.
   			  \and   			 
   			 Dipartimento di Fisica e Astronomia, Universit\'{a} degli Studi di Firenze, Via G. Sansone 1, I-50019 Sesto Fiorentino, Italy.
   			 \email{asa.skuladottir@unifi.it}
   			   \and
   			   INAF/Osservatorio Astrofisico di Arcetri, Largo E. Fermi 5, I-50125 Firenze, Italy.
   			   \and
   			  Department of Physics, Faculty of Science and Engineering, Konan University, 8-9-1 Okamoto, Kobe, Hyogo 658-8501, Japan.
   			   \and
				Monash Centre for Astrophysics, School of Physics and Astronomy, Monash University, VIC 3800, Australia.
                }

\abstract{
The slow ($s$) and intermediate ($i$) neutron ($n$) capture processes occur both in asymptotic giant branch (AGB) stars, and in massive stars. To study the build-up of the $s$- and $i$-products at low metallicity, we investigate the abundances of Y, Ba, La, Nd, and Eu in 98 stars, at $-2.4<\text{[Fe/H]}<-0.9$, in the Sculptor dwarf spheroidal galaxy. 
The chemical enrichment from AGB stars becomes apparent at $\text{[Fe/H]}\approx-2$ in Sculptor, and causes [Y/Ba], [La/Ba], [Nd/Ba] and [Eu/Ba] to decrease with metallicity, reaching subsolar values at the highest $\text{[Fe/H]}\approx-1$.
 To investigate individual nucleosynthetic sites, we compared three $n$-rich Sculptor stars with theoretical yields. One carbon-enhanced metal-poor (CEMP-no) star with high $\text{[Sr, Y, Zr]}>+0.7$ is best fit with a model of a rapidly-rotating massive star, the second (likely CH star) with the $i$-process, while the third has no satisfactory fit. For a more general understanding of the build-up of the heavy elements, we calculate for the first time the cumulative contribution of the $s$- and $i$-processes to the chemical enrichment in Sculptor, and compare with theoretical predictions. By correcting for the $r$-process, we derive $\text{[Y/Ba]}_{s/i}=-0.85\pm0.16$, $\text{[La/Ba]}_{s/i}=-0.49\pm0.17$, and $\text{[Nd/Ba]}_{s/i}=-0.48\pm0.12$, in the overall $s$- and/or $i$-process in Sculptor. These abundance ratios are within the range of those of CEMP stars in the Milky Way, which have either $s$- or $i$-process signatures. The low $\text{[Y/Ba]}_{s/i}$ and $\text{[La/Ba]}_{s/i}$ that we measure in Sculptor are inconsistent with them arising from the $s$-process only, but are more compatible with models of the $i$-process. Thus we conclude that both the $s$- and $i$-processes were important for the build-up of $n$-capture elements in the Sculptor dwarf spheroidal galaxy.

}

   \keywords{Stars: abundances --
                                Galaxies: dwarf galaxies --
                                Galaxies: individual (Sculptor dwarf spheroidal) --
                                Galaxies: abundances --
                                Galaxies: evolution
               }

   \maketitle

%
\section{Introduction} \label{sec:intro}

\subsection{The neutron-capture processes}

Three main neutron ($n$) capture processes are known to occur in environments characterized by different $n$-densities (low to high): the slow ($s$), the intermediate ($i$) and the rapid ($r$) process. The final chemical pattern created by each process depends on the $n$-density and the overall physical conditions at the production site (e.g. \citealt{Burbidge57,Cameron57,Sneden08}).

A generally accepted nucleosynthetic site for the main $s$-process are low- to intermediate-mass stars $(m_\star\lesssim8$~M$_\odot$) in the asymptotic giant branch (AGB) phase (e.g.~\citealt{Karakas14}). The heavy elements that are created by AGB stars are distributed to the surrounding interstellar medium (ISM) via stellar winds. This process therefore has a natural time delay relative to more massive stars, which depends on the lifetime of each star and when it reaches the AGB phase. The $s$-process also occurs in massive stars, mainly in their He-burning cores and C-burning shells, as the \textit{`weak'} s-process (e.g. \citealt{Langer89,Prantzos90,Raiteri91}). The efficiency of the weak $s$-process decreases with metallicity, and becomes almost zero below $Z\sim10^{-4}$ \citep{Prantzos90}. However, given sufficient $^{22}$Ne ($n$-source via the $^{22}$Ne($n$,
$\alpha$)\,$^{25}$Mg reaction), rotation can boost an enhanced $s$-process in massive stars (e.g. \citealt{Pignatari08,Chiappini11,Frischknecht12,Choplin18,Limongi18}). The impact of rotation on $s$-process yields in massive stars may be particularly important at low metallicity \citep{Frischknecht16}.

The $i$-process was originally proposed by \citet{CowanRose77}. Its abundance pattern overlaps with that of the $s$-process, but overall it is different from both the $s$- or $r$-processes, or a mixture of the products of the two (e.g. \citealt{Fishlock14,Hampel16,Roederer16}). The nucleosynthetic sites of the $i$-process remain unconfirmed  (e.g. \citealt{Frebel18,Koch19}), but among the proposed scenarios are: low-mass, low-metallicity ($\text{[Fe/H]}\lesssim-3$) stars \citep{Campbell08,Campbell10,Cruz13,Cristallo16}, massive ($5-10$~M$_\odot$) super-AGB stars \citep{Doherty15,Jones16}, evolved low-mass stars \citep{Herwig11,Hampel19}, and rapidly accreting white dwarfs \citep{Herwig14,Denissenkov17}. Finally, massive ($m_\star>20$~M$_\odot$), metal-poor stars could also play a role in the production of $i$-process elements \citep{Clarkson18,Banerjee18}.

The physical sites for the $r$-process are still strongly debated, and there might be more than one dominant source (e.g. \citealt{Cote19}). Among the suggested nucleosynthetic sites are: binary neutron star or black hole mergers \citep{Lattimer77,Rosswog99,Freiburghaus99,Watson19}, magnetohydrodynamically (MHD) driven supernovae (SN; \citealt{Winteler12,Nishimura15,Nishimura17}), collapsars \citep{Siegel19a,Siegel19b}, and core-collapse supernovae (ccSNe) \citep{Mathews90,Mathews92,Wheeler98,Ishimaru99,Ishimaru04,Ishimaru05,Wanajo03,Wanajo09}.

From an observational point of view, the large scatter of lighter $n$-capture elements (e.g. Sr, Y, Zr) at $\text{[Fe/H]}<-2.5$ in Milky Way halo stars, suggests that a process different from the main $s$-, $i$-, and $r$-processes is needed to explain the observations \citep{Travaglio04,Francois07,Hansen12}. The nucleosynthetic site of this process, the \textit{`Lighter-Element Primary Process'} (LEPP), also known as \textit{`weak'} or \textit{`limited'} $r$-process, is unconfirmed. This weak $r$-process might for example occur in neutrino-driven winds from ccSN \citep{Wanajo01,Wanajo11,Qian03,Montes07,Arcones11,Hansen14,Hirai19}. 

All these processes play a role in creating the various $n$-capture elements. For practical reasons, however, Ba is often considered the {`canonical'} tracer of the $s$-process, and Eu for the $r$-process. These elements are relatively easy to measure in stellar spectra, and $\sim$85\% of the solar Ba abundance comes from the $s$-process, while the $r$-process is responsible for $\sim$94\% of the Eu in the Sun \citep{Bisterzo14}. We note that the analysis of \citet{Bisterzo14} does not take the $i$-process into account, which is expected to create a significant amount of both Ba and Eu. However, since the $i$-process sources are still being debated, the overall contribution of the $i$-process to the chemical enrichment of the Milky Way is very uncertain.

\subsection{Carbon-enhanced metal-poor stars}

Carbon-enhanced metal-poor (CEMP) stars, which have $\text{[C/Fe]}>+0.7$ and $\text{[Fe/H]}\leq-2$, are categorized into CEMP-no/$s$/$i$/$r$ based on their heavy elemental abundance pattern (e.g. \citealt{Beers05,Aoki07,Hampel16}). The CEMP-$s$ stars (high Ba), have proven to be mostly in binary systems, and it is generally accepted that the overabundance of C and the heavier elements in most of these stars is caused by accretion from a companion AGB star \citep{Lucatello05,Starkenburg14,Abate15a,Abate15b,Abate18,HansenT16s}, while some might show the nucleosynthetic products of \textit{spinstars}, for instance, rapidly rotating (single) massive stars \citep{Choplin17}. Similarly, it has been suggested that the CEMP-$i$ stars (enhanced in both Ba and Eu; also known as CEMP-$r/s$ stars) are the result of a mass transfer from a binary companion that has undergone the $i$-process \citep{Lugaro09,Campbell10,Dardelet15,Hampel16,Koch19}. The CEMP-$r$ (high Eu) stars are believed to be polluted in heavy elements mainly by one source \citep{Frebel18}. These CEMP-$s/i/r$ stars are thus ideal for studying individual nucleosynthetic sites of the $s$-, $i$-, and $r$-processes. 

The so-called CEMP-no stars have no enhancements of $s$- or $r$-process elements (i.e. Ba and Eu). These stars have significantly lower binary fractions than CEMP-$s$ stars \citep{Starkenburg14,HansenT16no}. This, combined with the increasing fraction of these stars and more extreme carbon-enhancements towards the lowest [Fe/H] has lead to the belief that the CEMP-no stars show the imprints of the early generations of massive, extremely metal-poor stars (e.g. \citealt{UmedaNomoto03,MeynetMaeder06,Norris13,Salvadori15,deBennassuti17}). However, recent observations of CEMP-no stars indicate higher binary fraction compared to C-normal stars \citep{Arentsen19}. Furthermore, the general approach of measuring carbon through molecular bands, assuming 1D and local thermodynamic equilibrium (LTE), might significantly overestimate their C-abundances \citep{Amarsi19,AmarsiNS19,Norris19}. 

The CEMP-no stars in dwarf galaxies are not extensively studied (e.g.~\citealt{Salvadori15}). However, three CEMP-no stars in dwarf galaxies are of special interest: ET0097\footnote{This star shows some evidence of being a binary \citep{Skuladottir17}, however, binary transfer cannot easily explain the observed abundance pattern \citep{Skuladottir15a}.} in Sculptor \citep{Skuladottir15a}, ALW-8 in Carina \citep{Susmitha17}, and 10694 in Pisces~II \citep{Spite18}. These three stars all show enhancements of $\text{[Sr,Y,Zr/Ba]}>+0.7$, which is not typical for CEMP-no stars in the Milky Way (e.g. \citealt{Allen12}, and references therein). However, stars with high [Sr,Y,Zr/Ba] but $\text{[C/Fe]}<0$ have been observed in the Galactic halo \citep{Honda04,Honda06,Honda07} and in $\omega$~Cen \citep{Yong17}. Thus, it is unclear whether the enhancements of carbon and the lighter $n$-capture elements is connected to the same source, or specifically to the chemical evolution of dwarf galaxies, or if it is merely a coincidence. 

        \begin{figure*}
   \centering
   \includegraphics[width=\hsize-3cm]{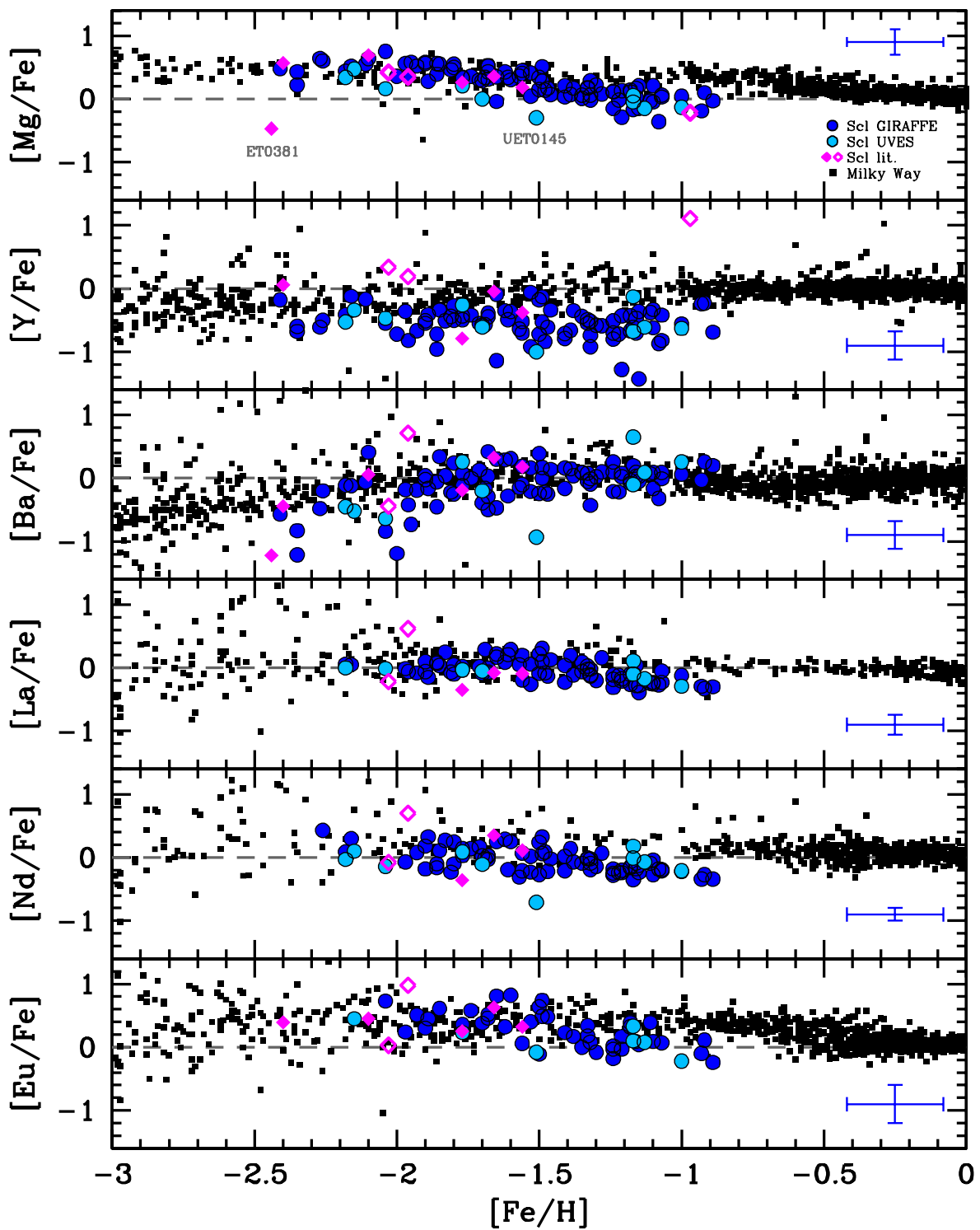}
      \caption{Abundances of Mg and $n$-capture elements in the Sculptor dSph. Target stars are shown with blue (GIRAFFE) and light-blue (UVES) circles. Previous measurements in Sculptor from HR spectra are shown with magenta diamonds. Open diamonds are stars with peculiar abundances of the $n$-capture elements. Milky Way stars are shown with black squares. \textit{Sculptor literature:} \citealt{Shetrone03,Geisler05,Kirby12,Skuladottir15a,Jablonka15}. \textit{Milky Way references}: \citealt{Burris00,Reddy03,Reddy06,Venn04,Simmerer04,Francois07,Hansen12,Mishenina13,Roederer14}. The SAGA database \citep{Suda08} was used to gather this compilation.
      }
         \label{fig:ncap}
   \end{figure*}

\subsubsection{The Sculptor dwarf spheroidal galaxy}

The first paper of this series, \textit{Neutron capture elements in dwarf galaxies~I: Chemical clocks and the short timescales of the $r$-process} (\citealt{Skuladottir19}; hereafter Paper~I), presented new measurements of Mg, Y, Ba, La, Nd, and Eu of 98 red giant branch (RGB) stars in the Sculptor dwarf spheriodal (Sph) galaxy from ESO VLT/FLAMES high-resolution (HR) spectra. In continuation of that work, here we investigate the heavy element pattern from the LEPP, $s$-, and $i$-processes, using the sample and chemical analysis from Paper~I. These stars have previously been carefully studied in the literature, regarding their light, $\alpha$- and iron peak elements \citep{Tolstoy09,North12,Skuladottir15b,Skuladottir17,Skuladottir18,Hill18}. 
The third paper in this series: \textit{Neutron capture elements in dwarf galaxies~III: A homogenized analysis of 13 dwarf spheroidal and ultra-faint galaxies} (Reichert et al., submitted to A\&A; hereafter Paper~III), presents a homogeneous analysis of $n$-capture elements in various dwarf galaxies.

The Sculptor dSph galaxy is a satellite of the Milky Way with stellar mass $\sim$10$^6$~M$_\odot$. Its stellar population is predominantly old, $>10$~Gyr \citep{deBoer12,Weisz14,Bettinelli19}, and metal-poor; the metallicity distribution function of stars in Sculptor peaks at $\text{[Fe/H]}\approx-2$ (e.g.~ \citealt{Battaglia08a}; see also Paper~I). The slow chemical enrichment of Sculptor, and other Milky Way satellite dwarf galaxies, provides a unique opportunity to investigate the cumulative contribution of metal-poor AGB stars, which otherwise can only be studied via their mass transfer onto a binary companion.


\section{Chemical abundance analysis}

A table with the chemical abundance measurements of Mg, Y, Ba, La, Nd, and Eu for 98 RGB stars in the Sculptor dSph is available via Paper~I. In total, 88~stars were observed with 4-5 settings of FLAMES/GIRAFFE, $R\sim20\,000$, and 10~stars with FLAMES/UVES, $R\sim47\,000$.

The Mg measurements were made using the \ion{Mg}{I} line at 5528.4~\AA, and two additional lines were also used in the UVES spectra, at 5183.6 and 5711.1~\AA\ (Paper~I). In our range of stellar parameters, $0\leq \log{g}\leq 1.3$, and $3800\leq T_\textsl{eff}/$K$\leq 4700$ \citep{Hill18}, the NLTE effects of the 5528.4~\AA\ line are typically small, $\sim$0.05~dex, with no significant metallicity dependence \citep{Bergemann15,NLTE_MPIA}.

In the GIRAFFE spectra, three lines were used for the abundance measurements of Y, at 4854.9, 4883.7 and 4900.1~\AA. These lines were visible in all spectra where the wavelength range was covered, and several additional lines were also used for the UVES spectra, giving overall consistent results. Two \ion{Ba}{II} lines, at 6141.7 and  6496.9~\AA\ were used for all spectra, and one additional line at 5853.7~\AA\ was used in the UVES spectra. 
A total of 8 \ion{La}{II} and 20 \ion{Nd}{II} lines were available in the GIRAFFE spectra, and approximately twice as many in the UVES spectra. However most of these lines become too weak to measure at $\text{[Fe/H]}\lesssim-2.2$. The same was true for the one \ion{Eu}{II} line that was used at 6645.1~\AA. For a full linelist and further description of the abundance determination see Paper~I.

The NLTE effects of Y, La, and Nd lines are not extensively studied, but the robustness of the $r$-process pattern in stars of different evolutionary stages (e.g.~\citealt{Sneden08}) suggests that the corrections for La and Nd are not severe. The NLTE effects of Ba in evolved RGB stars have proven to be minimal, $<0.1$~dex (e.g.~\citealt{Andrievsky09,Andrievsky17}), relative to typical abundance errors of dwarf galaxy stars. The assumption of LTE in this work therefore does not significantly affect our results or conclusions.

\section{Neutron-capture elements in Sculptor}

The chemical abundances of $n$-capture elements in the Sculptor dSph galaxy are shown in Fig.~\ref{fig:ncap}. To establish a reference for chemical enrichment timescales in Sculptor, we also include [Mg/Fe]. Mg traces ccSNe nearly exclusively, while Fe comes from both ccSNe and SN type~Ia (e.g.~\citealt{Tsujimoto95,Iwamoto99}). The [Mg/Fe] with [Fe/H] in Fig.~\ref{fig:ncap} follows a very well-known trend in Sculptor (e.g. \citealt{Geisler05,Tolstoy09,Kirby11,Hill18}). A~`knee' in [$\alpha$/Fe] is observed at $\text{[Fe/H]}\approx-1.8$ when SN type Ia start to significantly enrich the environment, $1-2$~Gyr after the onset of star formation \citep{deBoer12}. This happens at lower metallicity in the Sculptor dSph compared to the Milky Way, possibly due to lower star-formation rate and/or more metal loss (see a detailed discussion in \citealt{Hill18}).

Two stars in Sculptor, ET0381 \citep{Jablonka15} and UET0145 (this work; \citealt{Hill18}), have subsolar abundances in most [X/Fe], which is unusual at their metallicity, and are therefore specifically labeled in the top panel of Fig.~\ref{fig:ncap}. Furthermore, three stars in Sculptor with HR spectra \citep{Shetrone03,Geisler05,Skuladottir15a} have enhanced $n$-capture element abundances and are marked with open diamonds in Fig.~\ref{fig:ncap}. The [Mg/Fe] value for these three peculiar stars falls within the average values in Sculptor, and the same is true for other $\alpha$- as well as the iron peak elements \citep{Shetrone03,Geisler05,Skuladottir15a}. The following sub-sections will focus on the properties of the main stellar population in Sculptor, while these outliers are further discussed in Section~\ref{sec:pec}.

\subsection{The evolution of the $n$-capture elements} \label{sec:nevolution}

At the earliest times in the evolution of a galaxy, AGB stars have not started to contribute to the chemical evolution, and therefore Ba is mostly created by the $r$-process, with a possible contribution from massive, fast-rotating, low-metallicity stars (e.g. \citealt{Frischknecht16,Choplin18,Banerjee18}), and/or the $i$-process in massive stars (e.g. \citealt{Clarkson18}). In Fig.~\ref{fig:ncap}, we see very low $-1.5<\text{[Ba/Fe]}<0$ at $\text{[Fe/H]}<-2$, consistent with this scenario. 

As more intermediate- and low-mass stars go through the AGB phase, the $s$-process becomes more important. Thus [Ba/Fe] increases with increasing [Fe/H], reaching a plataeu at the solar value, $\text{[Ba/Fe]}\approx0$ at $\text{[Fe/H]}\gtrsim-1.6$, similar to the Milky Way. However, at these metallicities the contribution of Fe from SN type~Ia is much more significant in Sculptor compared to the Milky Way, as seen by the low [Mg/Fe] abundances (top panel of Fig.~\ref{fig:ncap}). Therefore, since [Ba/Fe] are similar in these galaxies, [Ba/Mg] is higher in Sculptor stars than in Milky Way stars at these metallicities (see Fig.~1 in Paper~I). We can thus conclude that the contribution of the $s$-process relative to ccSN is higher in Sculptor than in the Milky Way at the same $-1.6\leq\text{[Fe/H]}\leq-0.9$. Furthermore, the plateau of [Ba/Fe] in both Sculptor and the Milky Way suggests that the timescales of SN type~Ia and the $s$-process are comparable.

On the contrary, if we focus on the $r$-process element Eu, we notice high values, $\text{[Eu/Fe]}>0$ at the lowest metallicities, both in Sculptor and the Milky Way (bottom panel of Fig.~\ref{fig:ncap}). As the contribution of SN type~Ia to Fe increases, the [Eu/Fe] decreases, similar to the trend observed in [Mg/Fe] (top and bottom panels of Fig.~\ref{fig:ncap}). The same trend is observed in the Milky Way at higher metallicity, $\text{[Fe/H]}>-1$, when SN type~Ia start to dominate the chemical enrichment. The observed differences of Ba and Eu abundances in Sculptor are in good agreement with the generally accepted premises where the bulk of these elements come from distinct production sites, i.e., the $s$- and $r$-process, respectively. 

The elements La and Nd are predicted to be created by both the $s$- and $r$-processes, with 75\% and 58\%, respectively, of the solar abundance coming from the $s$-process \citep{Bisterzo14}. The chemical abundances of these elements in Sculptor bear signatures of this. Indeed, in Fig.~\ref{fig:ncap} we see that both [La/Fe] and [Nd/Fe] decrease with increasing [Fe/H]. However, the slope is not as steep as in the case of [Eu/Fe], since the delayed contribution of the $s$-process, somewhat counteracts that of SN type~Ia, but not fully as in the case of [Ba/Fe].

       \begin{figure}
   \centering
   \includegraphics[width=\hsize-1cm]{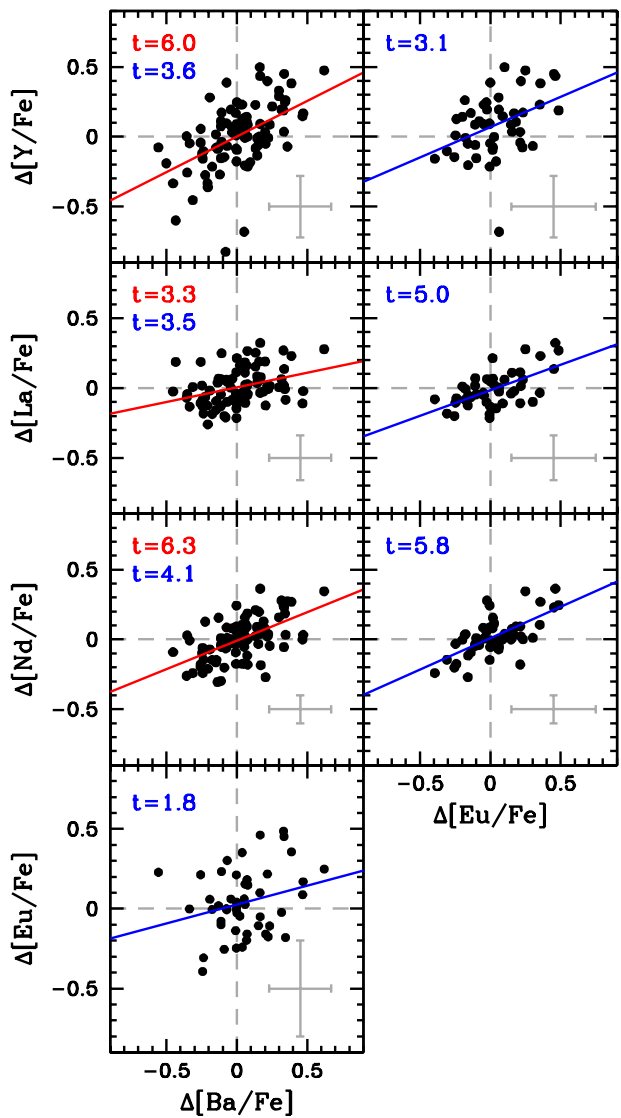}
      \caption{Correlations in scatter between different elements. The deviation from the mean abundance, $\Delta \text{[X/Fe]}$, for our target stars as a function of the same for Ba (left), and Eu (right). Lines show the best fit to the data. Red lines and t-values refer to all available data for Ba, blue to stars where Eu could be measured, i.e. blue t-values in left panels correspond to when the stellar sample on the right is used. Typical measurement errors are shown in gray.
      }
         \label{fig:corr}
   \end{figure}   

Finally, the light $n$-capture element Y is predicted to come from the LEPP, $s$-, $i$-, and $r$-processes \citep{Travaglio04,Sneden08,Karakas14,Hampel16}. According to \citet{Bisterzo14}, around 72\% of the solar Y abundance is expected to come from the $s$-process, which is comparable to La (75\%), however the behavior of Y and La in Sculptor is different, see top panel of Fig.~\ref{fig:ncap}. Furthermore, [Y/Fe] in Sculptor is different from the Milky Way, with significantly lower values, especially at higher metallicities ($\text{[Fe/H]}>-2$). Again, this is affected by SN type~Ia, but also by the lower Y yields in low-metallicity AGB stars compared to those at higher metallicities \citep{Karakas14}. The values of [Y/Fe] with increasing metallicity at $\text{[Fe/H]}>-2$ are consistent with a fairly flat or slightly decreasing trend, indicating that AGB stars are a significant source of this element in Sculptor, since it counteracts the effect of SN type~Ia (see also Paper~I). The production of Y before the onset of AGB stars ($\text{[Fe/H]}<-2$) is, however, less certain, but is further discussed in Section~\ref{sec:yba}.\\


\subsection{Scatter in the $n$-capture abundances} \label{sec:scatter}

In some cases, the observed scatter of [X/Fe] at a given [Fe/H] in Sculptor exceeds what is expected from the abundance measurement errors alone, see Fig.~\ref{fig:ncap} and representative error bars therein. This is particularly evident for [Ba/Fe] at low metallicities, $\text{[Fe/H]}\leq-1.8$, where Ba traces the $r$-process (see Section~\ref{sec:nevolution}). As the $r$-process products are known to come from strong and rare events, and the ISM is not fully mixed at the early stages (see discussion in Paper~I), a significant scatter is expected at the lowest metallicities, as is also observed in the Milky Way, see Fig.~\ref{fig:ncap}. Unfortunately, Eu, La, and Nd are challenging to measure at $\text{[Fe/H]}<-2$ with the available spectra in Sculptor, so low values would go undetected. We are therefore unable to confirm if this scatter is also present in these elements at low [Fe/H], but in the case of Y the scatter ($\sigma_\text{[Y/Fe]}=0.20$ at $\text{[Fe/H]}\leq-2$) is compatible with measurement errors ($\langle\delta_\text{[Y/Fe]}\rangle=0.21$).

At higher $\text{[Fe/H]}> -1.8$, the scatter in most elemental abundances is consistent with measurement errors. A possible exception is [Y/Fe] at $\text{[Fe/H]}>-1.3$, where $\sigma_\text{[Y/Fe]}=0.28$, while the average error is $\langle\delta_\text{[Y/Fe]}\rangle=0.23$. Another case is Nd, where the high number of available \ion{Nd}{II} lines leads to relatively small measurement errors. At intermediate metallicities, $-1.9\lesssim\text{[Fe/H]}\lesssim-1.3$, the scatter in [Nd/Fe] thus exceeds what is expected from the measurement errors alone, see Fig.~\ref{fig:ncap}.

An intrinsic scatter in these elements could be due to the $s$- and/or $r$-process products not being fully homogeneously mixed in the ISM. If the scatter is not solely from random errors, this would cause correlations in the abundances of different elements formed in the same process. For example, a star with a high $s$-process contribution for its [Fe/H] would have high abundances of several $n$-capture elements, $\text{[X/Fe]}>\langle\text{[X/Fe]}\rangle$. To test this, we determined a running mean for each element, $\langle \text{[X/Fe]}\rangle$, as a function of [Fe/H] and for each star we calculated the deviation from this mean, $\Delta \text{[X/Fe]}=\text{[X/Fe]}-\langle\text{[X/Fe]}\rangle$. To avoid systematic offsets between surveys, we only include our current sample. The clear outlier UET0145 (see Fig.~\ref{fig:ncap}) is also excluded from these results and the following discussion.\footnote{Including UET0145 would make our measured correlations stronger.}

The correlations between the scatter of different elements with $\Delta \text{[Ba/Fe]}$ and $\Delta \text{[Eu/Fe]}$ are shown in Fig.~\ref{fig:corr}, with the corresponding t-values, $t=r\sqrt{N-2}/\sqrt{1-r^2}$, where $N$ is the sample size and $r$ is the Pearson correlation coefficient. These $t$-values measure the significance of the correlation, where the 98\% confidence level is at $t \approx2.4$ for our sample sizes (50-100 stars). Due to fewer available measurements, any correlations with Eu include only $\approx~50$~stars while those with Ba typically have $>80$~stars (see Paper~I). The level of significance of these correlations is directly affected by the differences in sample sizes and average data quality. (The Eu sample is biased towards stars with higher S/N spectra and $\text{[Fe/H]}\gtrsim-2$; Paper~I). For a fairer comparison, we therefore also list the $t$-values for the reduced sample for Ba (blue in Fig.~\ref{fig:corr}), only including stars which have a measurement in both Ba and Eu. 

A significant correlation of $\Delta \text{[Ba/Fe]}$ with both $\Delta \text{[Y/Fe]}$ and $\Delta \text{[Nd/Fe]}$ is seen in Fig.~\ref{fig:corr}, while the correlation with $\Delta \text{[La/Fe]}$ is less strong. This suggests that the $s$-process products are not fully homogeneously mixed in the ISM of Sculptor. This is not surprising for the products of AGB stars, which are released by stellar winds that are much less energetic than SNe. Similarly, $\Delta \text{[Eu/Fe]}$ correlates clearly with $\Delta \text{[La/Fe]}$ and $\Delta \text{[Nd/Fe]}$; and, to a lesser degree, also with $\Delta \text{[Y/Fe]}$. These results indicate inhomogeneously mixed $r$-process products in the ISM of Sculptor, even at the higher metallicities $\text{[Fe/H]}\gtrsim-2$ where we are able to measure Eu. This is likely caused by the rarity of $r$-process events.

We note that $\Delta \text{[Eu/Fe]}$ shows a more significant correlation with $\Delta \text{[Nd/Fe]}$ compared to $\Delta \text{[La/Fe]}$. Although Nd is expected to have a stronger contribution from the $r$-process compared to La, most likely the smaller errors of Nd result in a more significant correlation. Indeed, the correlation of $\Delta \text{[Ba/Fe]}$ is also stronger with Nd than La, though the $s$-process is expected to be slightly more effective in producing La (e.g.~\citealt{Bisterzo14}). However, the correlation of Eu to both of those elements is stronger than with Ba, suggesting that the intrinsic scatter in Nd and La might be more strongly affected by the $r$-process than the $s$-process. 

Although not shown in Fig.~\ref{fig:corr}, there is also a statistically significant correlation of the scatter in Nd with that of Y and La, with $t$-values of 4.1 and 5.6 respectively, when including all available measurements. This is expected, as all these elements are created both by the $s$- and $r$-processes.

Finally, we note that some part of our correlations might arise from systematic errors, such as stellar parameters and continuum evaluation of the spectra. However, in agreement with their different origin, there is no statistically significant correlation between the scatter in Ba and Eu (bottom panel of Fig.~\ref{fig:corr}). If Ba traces the $r$-process at low metallicity, some correlation would be expected with Eu, but given our limited number of stars with Eu measurements at $\text{[Fe/H]}<-2$ (and the relatively high measurement errors, see Fig.~\ref{fig:ncap}, and \ref{fig:corr}), it is unsurprising that the correlation is not stronger. Furthermore, no clear correlation was found between the scatter in Mg and any of the $n$-capture elements.

Ultimately, the correlations shown in Fig.~\ref{fig:corr} indicate that some of the observed scatter in Y, Ba, La, Nd, and Eu is intrinsic. The products of both the $s$- and $r$-process were therefore likely well but not completely homogeneously mixed in the ISM during the evolution of the Sculptor dSph.

\section{The relative contribution of different $n$-capture processes in Sculptor} \label{sec:ba}

             \begin{figure}
   \centering
   \includegraphics[width=\hsize-0cm]{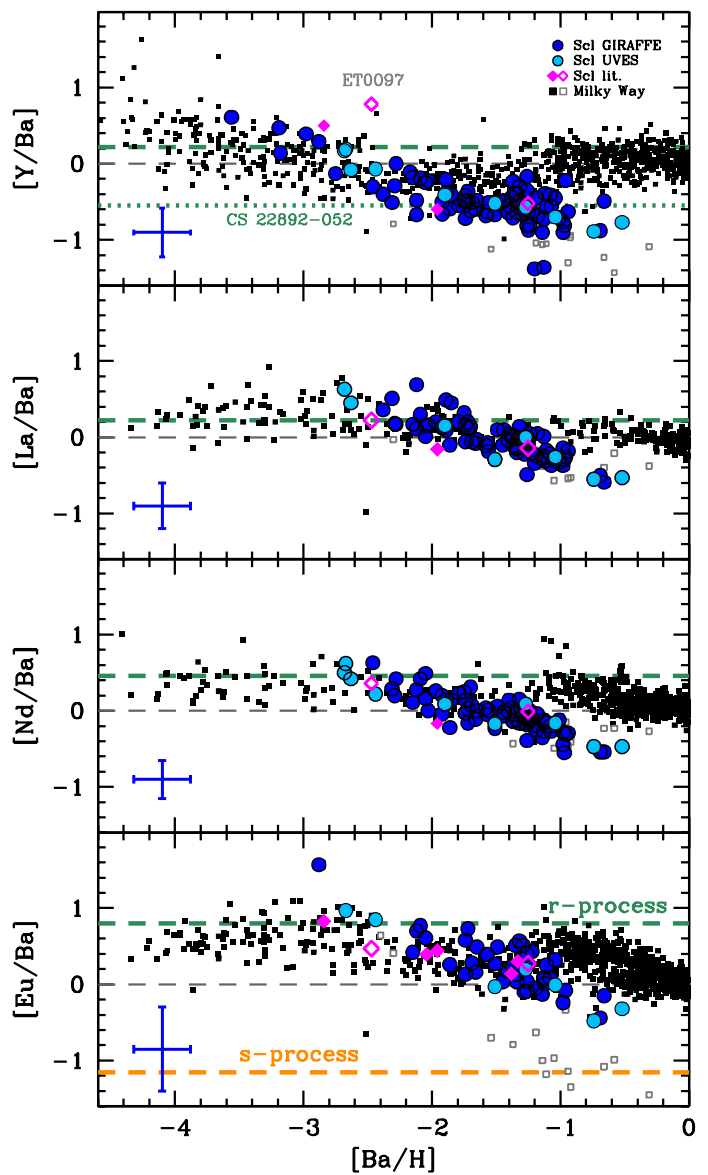}
      \caption{Ratios of $n$-capture elements to Ba as a function of [Ba/H]. Symbols and Sculptor references are the same as in Fig.~\ref{fig:ncap}, but open squares are confirmed Milky Way CEMP stars with $\text{[Ba/Mg]}>0$. The [Eu/Ba] ratio of the pure $r$-process is shown with a green dashed line, and the pure $s$-process is in orange \citep{Bisterzo14}. In the case of [Y/Ba] the value of the $r$-rich star CS~22892-052 is also shown with a green dotted line
      \citep{Sneden03}. \textit{Milky Way references}: \citealt{Reddy03,Reddy06,Venn04,Francois07,Hansen12,Mishenina13,Roederer14}.
      }
         \label{fig:ncapba}
   \end{figure}   
   
 \begin{figure}
   \centering
   \includegraphics[width=\hsize-1.5cm]{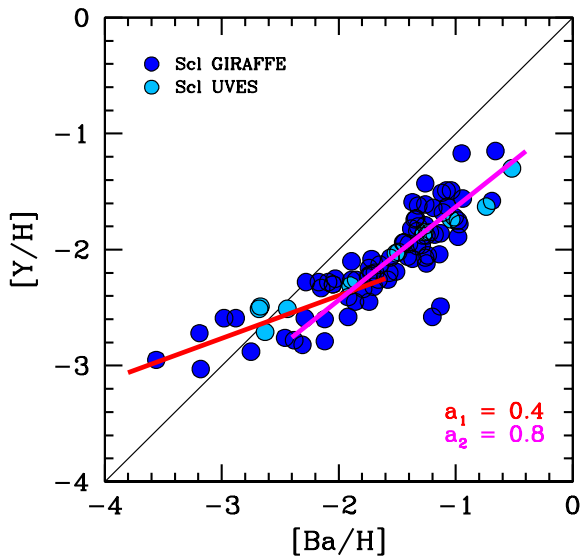}
      \caption{Abundances of Y and Ba in Sculptor. The red line with a slope of $a_1=0.4$ is the best fit for $\text{[Ba/H]}<-2.1$, while the magenta line with slope $a_2=0.8$ is the best fit for data with $\text{[Ba/H]}>-1.9$. The two low outliers in [Y/H] at $\text{[Ba/H]}\approx-1$ are not included in the fit. Solid black line shows $\text{[Y/Ba]}=0$.
      }
         \label{fig:logeps}
   \end{figure}   
   
\subsection{Chemical evolution with Ba}   

The relative contribution of different processes to each $n$-capture element in the Sculptor dSph can be examined by plotting Y, La, Nd, and Eu over Ba against [Ba/H], see Fig.~\ref{fig:ncapba}. At the earliest times, the production of Ba is dominated by the $r$-process, as the low- to intermediate-mass stars have yet to reach their AGB phase (see Section~\ref{sec:nevolution}). The bottom panel of Fig.~\ref{fig:ncapba} confirms this, with high [Eu/Ba] ratios in Sculptor at $\text{[Ba/H]}<-2$, consistent with the pure (or almost pure) $r$-process. At these lowest $\text{[Ba/H]}<-2$ the heavy $n$-capture elements (La, Nd, and Eu) all reach supersolar ratios $\text{[X/Ba]}>0$, comparable to those of the Milky Way at the same [Ba/H], where the heavy element enrichment is dominated by the $r$-process. 

At $\text{[Ba/H]}>-2$, the influence of AGB stars becomes apparent in Fig.~\ref{fig:ncapba}, and [X/Ba] decreases with increasing [Ba/H] for all observed $n$-capture elements in Sculptor. This indicates that AGB stars in Sculptor have the strongest influence on Ba. Since subsolar values are reached at the end of the chemical evolution in Sculptor (Fig.~\ref{fig:ncapba}), it can be immediately deducted that $\text{[Y,La,Nd/Ba]}<0$ in the cumulative AGB enrichment at low metallicities, meaning in the integrated yields over all stellar masses (see further discussion in Section~\ref{sec:si}). At the highest $\text{[Ba/H]}\gtrsim-1.4$ , [X/Ba] in Sculptor are always lower than those of the main stellar population in the Milky Way (filled black squares in Fig.~\ref{fig:ncapba}). This suggests that the relative $s$-process (and/or $i$-process) to $r$-process contribution is higher in Sculptor at these [Ba/H]. This is confirmed by the lowest panel of Fig.~\ref{fig:ncapba}, where $\text{[Eu/Ba]}_\text{Scl}<\text{[Eu/Ba]}_\text{MW}$ at the highest metallicity, as expected by lower relative $r$-process contribution.

The different [X/Ba] trends in Sculptor and the Milky Way can be understood by the distinct star-formation histories of these two galaxies. With the Milky Way's higher star-formation rate, the gas is enriched up to higher metallicities (including [Ba/H]) before the production by low- and intermediate-mass AGB stars becomes significant. Thus, the [Eu/Ba] ratios start decreasing at higher [Ba/H] in the Milky Way compared to Sculptor, analogous to what is seen with [$\alpha$/Fe] due to the delayed contribution of SN type~Ia (Fig.~\ref{fig:ncap}; also \citealt{Hill18}). Contrary to the Milky Way, Sculptor's star formation died out\linebreak $\sim8-10$~Gyr ago \citep{deBoer12,Weisz14,Savino18,Bettinelli19}, presumably because the bulk of its gas was lost. This could further increase the impact of AGB yields at the later stages of the galaxy's evolution, if their products were released into a relatively small mass of gas (e.g.~\citealt{Salvadori15}).

Finally, if we focus on the Milky Way CEMP-$s$, and -$i$ stars in Fig.~\ref{fig:ncapba} (open squares), we see that [La/Ba] and [Nd/Ba] are compatible with the Sculptor values, while [Y/Ba] is on average lower in the Milky Way CEMP stars. In general the high-[Ba/H] Sculptor stars have more contribution from the $r$-process compared to Milky Way CEMP-$s$ and -$i$ stars. This can be seen in the higher [Eu/Ba] in the Sculptor stars compared to Milky Way CEMP-$s$ and -$i$ stars. In many cases, the latter have [Eu/Ba] values consistent with pure s-process (Fig.~\ref{fig:ncapba}). Stars in Sculptor which are peculiar in the $n$-capture elements are identified with open magenta diamonds in Fig.~\ref{fig:ncap} and~\ref{fig:ncapba}. These stars are discussed in Section~\ref{sec:pec}.

                    \begin{figure*}
   \centering
   \includegraphics[width=\hsize-0.3cm]{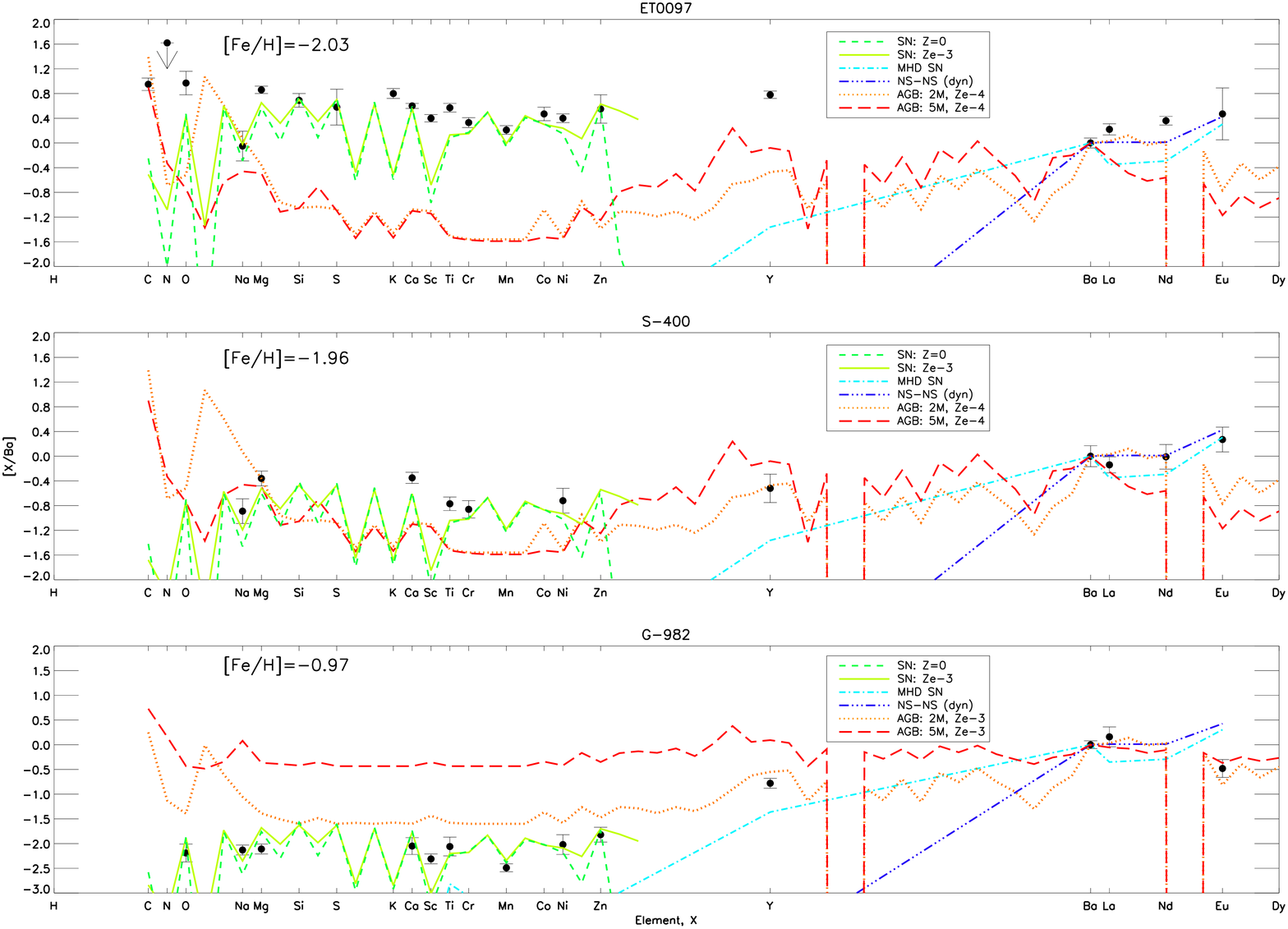}
      \caption{Black points are abundances of three peculiar stars in the Sculptor dSph (open diamonds in Fig.~\ref{fig:ncap} and \ref{fig:ncapba}): ET0097 at $\text{[Fe/H]}=-2$ \citep{Skuladottir15a}, Scl-400 at $\text{[Fe/H]}=-2$ \citep{Shetrone03}, and G-982 at $\text{[Fe/H]}=-1$ \citep{Geisler05}. Two AGB models ($Z=10^{-4}$) from the F.R.U.I.T.Y database are shown: 2~M$_\odot$ (salmon), and 5~M$_\odot$ (red). Two $r$-process models are included: MHD SN (cyan) from \citet{Winteler12}, and dynamic ejecta from NSM (blue) from \citet{Korobkin12,Rosswog13}. Models for ccSN are shown with green lines (dashed:~$Z=0$; solid:~$Z=10^{-3}$), using IMF-weighted yields from \citet{Kobayashi06}.
      }
         \label{fig:pecstars}
   \end{figure*}

\subsection{The relative production of Y and Ba} \label{sec:yba}

The ratio of [Y/Ba] in Sculptor follows the upper envelope of those observed in the Milky Way at $\text{[Ba/H]}<-2.4$ (top panel of Fig.~\ref{fig:ncapba}). This is in good agreement with the observational findings of \citet{Mashonkina17} who investigated the ratio of [Sr/Ba] in seven dwarf galaxies\footnote{Fornax, Sculptor, Sextans, Ursa Minor, Ursa Major~II, Bo{\"o}tes~I, and Leo~IV} at low metallicities, $\text{[Ba/H]}<-2$. Both Sr and Y are light $n$-capture elements belonging to the first peak, which have been shown as having very tight correlations (e.g. \citealt{Francois07,Hansen12}) and are generally accepted to have the same nucleosynthetic origin (e.g. \citealt{Sneden08}). \citet{Mashonkina17} found that dSph galaxies typically have increasing [Sr/Ba] ratios towards lower [Ba/H], similar to the higher values in the Milky Way, while the smaller UFD galaxies have lower ratios, $\text{[Sr/Ba]}\approx0$ (with significant scatter). Our new measurements of [Y/Ba] in Sculptor (Fig.~\ref{fig:ncapba}) are in very good agreement with these results (see also Paper~III).

Although the abundance pattern of the $r$-process has been shown to be robust at $55<Z<70$, the lighter $n$-capture elements, such as Y, show a large scatter in its ratios (e.g.~\citealt{Roederer10}). In Fig.~\ref{fig:ncapba} we show [Y/Ba] both for the solar-scaled $r$-process \citep{Bisterzo14}, but also for the $r$-rich star  CS~22892-052 at $\text{[Fe/H]}=-3$ \citep{Sneden03}. This demonstrates that the solar-scaled $r$-process is not adequate to describe [Y/Ba] in the $r$-process at low metallicities. From the decreasing values of [Y/Ba] with increasing pollution from metal-poor AGB stars (Fig.~\ref{fig:ncapba}), we infer that their yields have $\text{[Y/Ba]}<0$. Thus it is hard to explain the high $\text{[Y/Ba]}\gtrsim+0.4$ at the lowest [Fe/H], both in Sculptor and in the Milky Way (e.g.~\citealt{McWilliam98}). A significant contribution from another process such as the weak $r$-process (or LEPP) is therefore needed (e.g. \citealt{Francois07,Hansen12}). 

In Fig.~\ref{fig:logeps} we can see that there is a clear change in the slope of [Y/H] with [Ba/H] when the contribution of AGB stars starts to become significant, that is, around $\text{[Ba/H]} \approx-2$ (corresponding to $\text{[Fe/H]}\approx-1.8$). The slopes of [Y/H] with [Ba/H] in Fig.~\ref{fig:logeps} are: $a_1=0.4\pm0.1$ below $\text{[Ba/H]}<-2$; and $a_2=0.8\pm0.1$ for $\text{[Ba/H]}>-2$. Here the errors include: statistical error of the fit; whether outliers are included or not (see Fig.~\ref{fig:logeps}); and the effect of shifting the [Ba/H] boundary by $\pm0.2$~dex. 

If Y and Ba are completely co-produced in a single nucleosynthetic site, a slope between their abundances of $a=1$ is expected.\footnote{This holds assuming no metallicity nor stellar mass dependence of their ratios in the yields.} At higher [Ba/H] and thus later times, Y is therefore consistent with being mostly produced by the main $s$-process in Sculptor, while at the earliest times the production of Y and Ba is more likely from separate sources.

   \begin{figure*}
   \centering
       \includegraphics[scale=0.65]{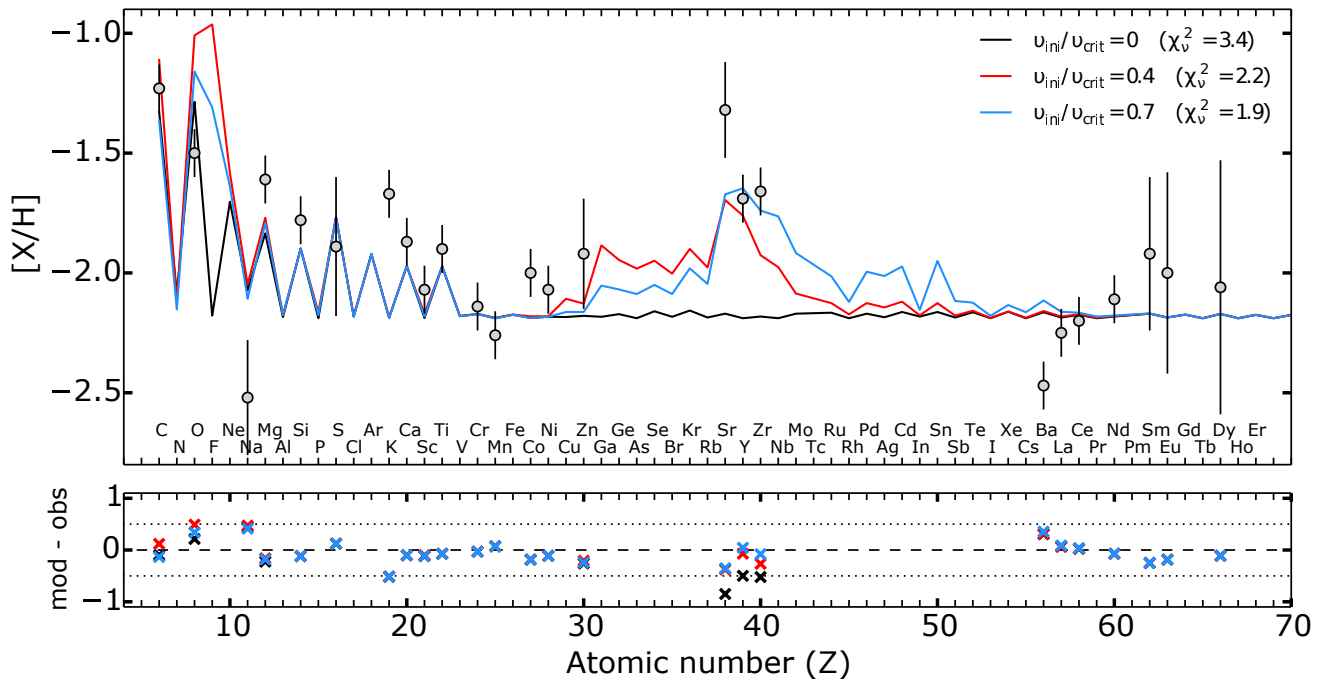}
   \caption{
Best fit massive-star models for the CEMP-no star ET0097 at $\text{[Fe/H]}=-2$. Points with errorbars are 1D LTE abundance measurements from \citet{Skuladottir15a}. Three models are shown with lines, computed at three different initial rotational velocities, $\upsilon_{\rm ini}/ \upsilon_{\rm crit} =$0 (black), 0.4 (red) and 0.7 (blue). All models are for 25~M$_\odot$ stars at $Z=10^{-3}$. The reduced $\chi^2$ of each fit are shown on the top right, and the residuals of the fits are in the bottom panel.
   }
\label{fig:rrms}
    \end{figure*}

\section{Comparison with theory - peculiar stars} \label{sec:pec}

The small abundance dispersion of the bulk of the stellar population in the Sculptor dSph may indicate that most of these stars were formed from a rather well-mixed reservoir enriched by a significant number of previous stellar generations (see Fig.~\ref{fig:ncap}, and \citealt{Hill18}). However, three previously published stars, ET0097, G-982, and S-400 are chemically peculiar, showing significant differences in the heavy element abundances compared to the main population, see Fig.~\ref{fig:ncap} and~\ref{fig:ncapba}. At these metallicities, $-2\leq\text{[Fe/H]}\leq-1$, it is unlikely that a single source is responsible for the complete enrichment of these stars (e.g. \citealt{Salvadori19}). However, the high $n$-capture elements suggest that the peculiarity of the abundance pattern of these stars at $Z>30$ might be the cause of binary transfer or a rare event, dominating the heavy element enrichment. Such peculiarities may thus indicate that the heavy elements in these stars mainly reflect the material ejected by just one or very few sources.

The detailed abundance patterns of the three peculiar stars are shown in Fig.~\ref{fig:pecstars}, along with comparison of models of AGB stars \citep{Cristallo15}, MHD SN \citep{Winteler12}, and dynamical ejecta from NSM \citep{Korobkin12,Rosswog13}. Contrary to the heavy elements, the $\alpha$- and iron peak element abundance patterns of these stars are typical for their metallicity (see e.g. [Mg/Fe] in Fig.~\ref{fig:ncap}, and \citealt{Hill18}). Thus it is likely that the $\alpha$- and iron peak elements arise from many generations of ccSN. For comparison, we therefore also include ccSN models in Fig.~\ref{fig:pecstars}, which have been weighted with the initial mass function (IMF) of stars, a power-law mass spectrum, with a slope of $x=1.35$, lower mass limit $M_l=0.07$~M$_\odot$, and upper mass $M_u=50$~M$_\odot$
\citep{Kobayashi06}.

We also compared the observed abundances of these peculiar stars with the undiluted $i$-process 
\citep{Hampel16}, and the ejecta of 25~M$_\odot$ massive star models computed at $Z=10^{-3}$ with\footnote{The critical velocity $\upsilon_{\rm crit}$ is reached when the gravitational acceleration is counterbalanced by the centrifugal force. It is expressed as $\upsilon_{\rm crit} = \sqrt{\frac{2}{3}\frac{GM}{R_{\rm p,c}}}$ with $R_{\rm p,c}$ the polar radius at the critical limit.} $\upsilon_{\rm ini}/ \upsilon_{\rm crit} =$0, 0.4 and 0.7, where the $s$-process is followed consistently during the evolution (details on the models can be found in \citealt{Choplin18}). To make the fits, the mass cut\footnote{Mass coordinate that delimitates the part of the star which is expelled from the part which is kept into the remnant.}, $M_{\rm cut}$, and the dilution factor\footnote{$D = M_{\rm ISM} / M_{\rm ej}$ where $M_{\rm ISM}$ is the mass of interstellar medium added to the ejected mass $M_{\rm ej}$.}, D, and the mass of added hydrogen are left as free parameters.

\subsection*{ET0097}

The star ET0097 is a CEMP-no star at $\text{[Fe/H]}=-2$, with unusually high $\text{[Sr,Y,Zr/Ba]}>+0.7$ \citep{Skuladottir15a}, but its [X/Ba] are otherwise in agreement with Sculptor stars, see Fig~\ref{fig:ncapba}. It has experienced internal mixing on the RGB \citep{Skuladottir15a} and the current C abundance is therefore smaller than the initial abundances, and the N abundance is higher \citep{Gratton00,Spite05,Placco14}. For C, we consider a correction of $+0.3$ dex, which gives [C/Fe] $=0.8$, and we assume the measured value of [N/Fe] is an upper limit.

The comparison of the chemical abundances of ET0097 with AGB and $r$-process models is shown in the top panel of Fig.~\ref{fig:pecstars}. The abundance pattern of the heavy $n$-capture elements ($Z>55$) are best fit with a NSM. However, the high [Y/Ba] abundance cannot be fitted with either the $r$-process or AGB model predictions. One-zone models of the pure $i$-process are also unable to reproduce the high [Eu/Ba] and [Sr,Y,Zr/Ba] \citep{Hampel16}.

Fig.~\ref{fig:rrms} shows the best fits for ET0097 while considering three rotating 25~M$_\odot$ stellar models. The reduced chi-square is computed as $\chi_{\nu}^2 = \chi^2 / (N-m)$, with $N$ the number of data points and $m$ the number of free parameters. The rotating models provide better solutions than the nonrotating model, which produce very small amounts of $n$-capture elements. However, there are some significant discrepancies even for the rotating models. The relatively high [Sr/H] ratio cannot be well reproduced together with the lower [Y/H] and [Zr/H] ratios. Also, the increasing trend from Ba to Nd cannot be reproduced. In particular, $\text{[Ba/Nd]}< 0$ cannot be obtained in the massive star models considered here. 

However, the analysis of \citet{Skuladottir15a} assumes 1D LTE, while some elements have quite strong NLTE effects at low metallicity, in particular Na and K. \citet{Skuladottir15a} evaluated Na from three lines at 5890.0, 5895.9, and 8183.3~\AA, where the two bluer lines have been shown to need extremely large positive corrections, $\lesssim+0.5$ \citep{Andrievsky07}. Similarly, the K resonance lines at 7664.9 and  7699.0~\AA, which are used in \citet{Skuladottir15a}, need large negative corrections \citep{Andrievsky10,Reggiani19}. The Sr in \citet{Skuladottir15a} is based on two available lines at 4077.7 and 4607.3~\AA, where the bluer line gives abundance 0.26~dex lower than the other. Dedicated studies of NLTE effects of Sr (e.g.~\citealt{Andrievsky11,Bergemann12}) usually do not include stellar parameters compatible to ET0097, i.e. $T_\textsl{eff}=4383$~K and $\log{g}=0.75$. However, \citet{Mashonkina17} investigate NLTE effects on the \ion{Sr}{I} 4077.7~\AA\ line in cool RGB stars in dwarf galaxies, and find that $\Delta\text{[Sr/H]}_\text{NLTE}\approx 0$ around $\text{[Fe/H]}=-2$ (see their Fig.~1). Given the overall uncertainties of the abundance analysis, ET0097 is therefore in fairly good agreement with being the descendant of rapidly-rotating, massive, low-metallicity star.

\subsection*{S-400}

The outlier Scl-400 \citep{Shetrone03} has similar [X/Ba] as other stars in Sculptor (see Fig.~\ref{fig:ncapba}), although it has a very high $\text{[Ba/H]}=-1.3$ for its $\text{[Fe/H]}=-2$ (see Fig.~\ref{fig:ncap}). The C-abundance of this star is unknown, but if the carbon is high, this would be a CEMP-$r$ star, since $\text{[Eu/Ba]}=+0.3$. In fact, it can be seen in the middle panel of Fig.~\ref{fig:pecstars} that the heavy elemental abundance pattern of S-400 cannot be well fit with the $s$-process in AGB stars, since [Eu/Ba] is too high. The heavier $n$-capture elements (Ba, La, Nd and Eu) are well fit with the $r$-process (both MHD SN and NSM), but it does not explain the high [Y/Ba] observed in this star. 

The abundances of S-400 are rising from Sr to Eu (with the exception of La). This trend cannot be reproduced by the massive star models considered here, which predict a flat or decreasing trend. Models of the pure $i$-process, are also unable to reproduce this pattern, since $\text{[Eu/Ba]}>0$.

\subsection*{G-982}

                \begin{figure}
   \centering
   \includegraphics[width=\hsize-0cm]{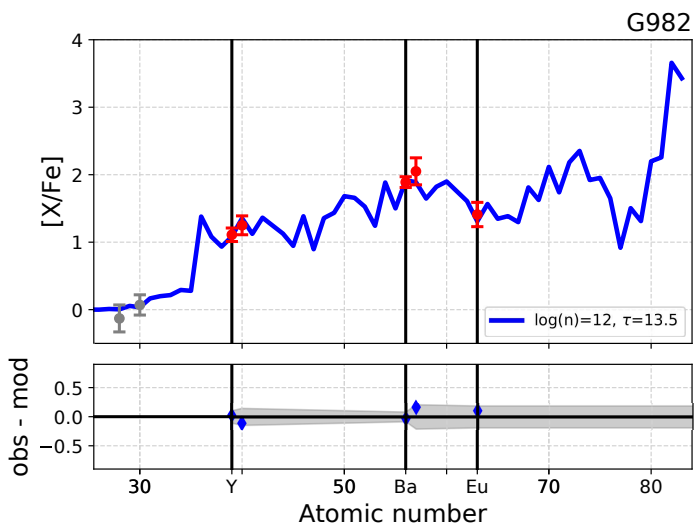}
      \caption{\textit{Top panel:} Best fitting $i$-process model for the Sculptor (likely CH) star G-982 at $\text{[Fe/H]}=-1$. Red points with error bars are abundance measurements from \citet{Geisler05}, and the $i$-process model from 
      \citet{Hampel16} is shown with a blue line. \textit{Bottom panel:} Residuals of the fit.}
         \label{fig:g982}
   \end{figure}

The star G-982 is metal-rich, $\text{[Fe/H]}=-1$, with very high values of the $n$-capture elements: $\text{[Y/Fe]}=+1.1$, $\text{[Zr/Fe]}=+1.3$, $\text{[Ba/Fe]}=+1.9$, $\text{[La/Fe]}=+2.1$, and $\text{[Eu/Fe]}=+1.4$ \citep{Geisler05}, and it is thus in most cases outside the range depicted in Figs~\ref{fig:ncap} and \ref{fig:ncapba}. \citet{Geisler05} did not provide C abundance for this star, but it was excluded from the analysis of \citet{Hill18}, under the name ET0136, due to strong molecular CN-bands. This is therefore likely a CEMP-$s$ or a CEMP-$i$ star (typically called CH stars at this metallicity). According to \citet{Geisler05}, G-982 has $\text{[Y/Ba]}=-0.8$, $\text{[La/Ba]}=+0.2$ and $\text{[Eu/Ba]}=-0.5$. Although the Y abundance is consistent with the most Ba-rich stars in Sculptor (see Fig.~\ref{fig:ncapba}), the La abundance is significantly higher, suggesting that G-982 was enriched by an atypical companion.

The extremely high Ba and La abundances make this star difficult to fit, and neither the AGB models (Fig.~\ref{fig:pecstars}) nor massive star models are successful. Although the abundance pattern of a 2~M$_\odot$ AGB star matches well with the heavy elements, it produces too much of the lighter elements ($Z<30$) to provide a reasonable fit. 
Similarly to S-400, the high [Ba/Fe] and [La/Fe] ratios of this star cannot be reproduced together with the low [Y/Fe] ratio in the massive star models from \citet{Choplin18}. 

One-zone models of the pure $i$-process are, on the other hand, able to well fit the abundance pattern of G-982, see Fig.~\ref{fig:g982}. However, this is based on only five measurements of heavy elements ($Z>30$). In particular, the model predicts extremely high $\text{[Pd/Fe]}=4$, which makes it unclear whether it is plausible.

                \begin{figure}
   \centering
   \includegraphics[width=\hsize-0cm]{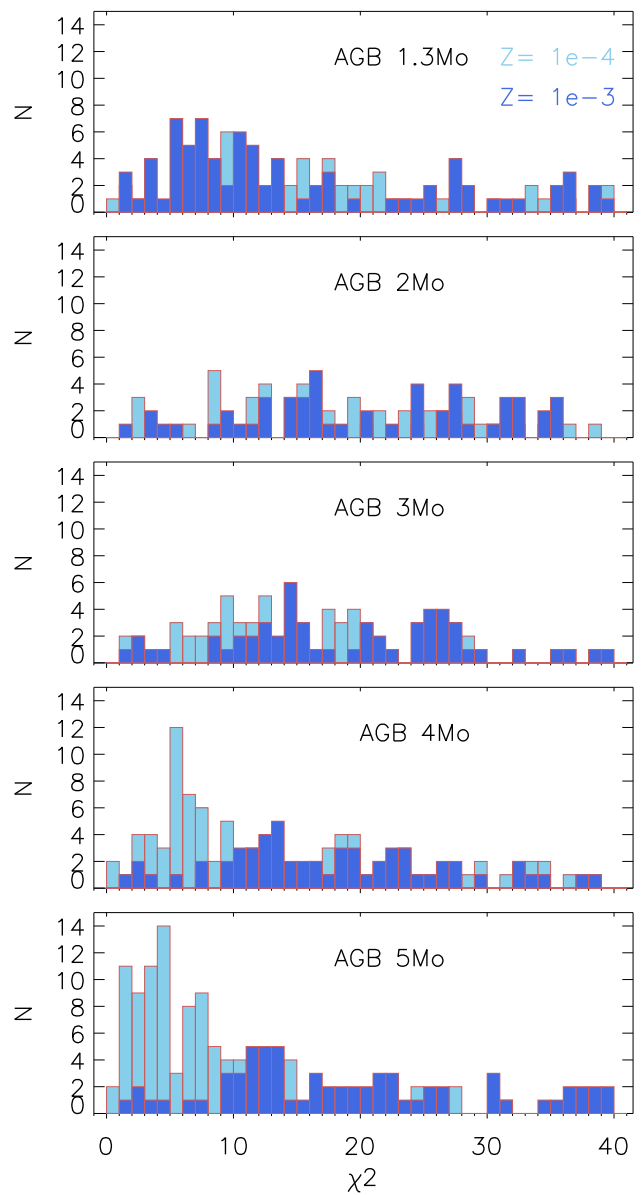}
      \caption{Number of stars, $N$, with a given goodness of fit, $\chi^2$, between the measured abundance pattern (Y, Ba, La, Nd, and Eu) of individual stars in Sculptor, and theoretical $s$-process AGB yields from the F.R.U.I.T.Y database. Results are shown for five different AGB masses: 1.3, 2, 3, 4 and 5~M$_\odot$ (from top to bottom), for two metallicities: $Z=10^{-4}$ (light blue) and $Z=10^{-3}$ (dark blue). Poor fits, $\chi^2>40$, are not shown.
      }
         \label{fig:histo}
   \end{figure}         

                \begin{figure}
   \centering
   \includegraphics[width=\hsize-0cm]{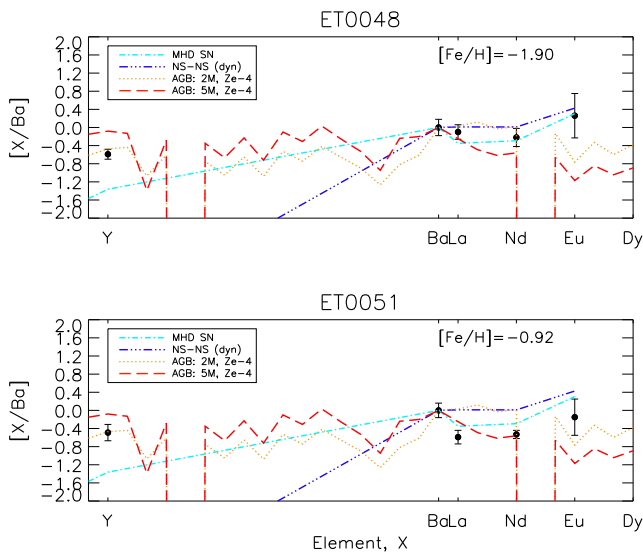}
      \caption{Abundances of two typical Sculptor stars are shown with black circles and error bars: ET0048 at $\text{[Fe/H]}=-1.90$, and ET0051 at $\text{[Fe/H]}=-0.92$. Two AGB models ($Z=10^{-4}$) from the F.R.U.I.T.Y database are shown: 2~M$_\odot$ (salmon), and 5~M$_\odot$ (red). Two $r$-process models are included: MHD SN (cyan) from \citet{Winteler12}, and NSM (blue) from \citet{Korobkin12,Rosswog13}.
      }
         \label{fig:twostars}
   \end{figure}

\section{Comparison with theory - main population}   

\subsection{Direct comparison}  \label{sec:yieldcomp}

The main stellar population in Sculptor formed out of an ISM, which was presumably enriched slowly over time. Each star is therefore expected to bear the signatures of a range of $n$-capture events, where the contribution of AGB stars became more dominant with increasing metallicity, corresponding to later times (see Section~\ref{sec:ba}; and Paper~I). However, it is still useful to investigate whether this contribution is dominated by AGB stars with a specific mass. Fig.~\ref{fig:histo} shows how well our observed abundance pattern for the heavy elements (Y, Ba, La, Nd, and Eu) is fit by different $s$-process AGB yields from the F.R.U.I.T.Y\footnote{http://fruity.oa-teramo.inaf.it/} database \citep{Cristallo09,Cristallo11}. The models in Fig.~\ref{fig:histo} assume a standard $^{13}$C-pocket, nonrotating AGB stellar models with masses between 1.3 and 5~M$_{\odot}$. We use models with metallicities that agree with the Sculptor stellar sample: $Z=10^{-4}$ and $Z=10^{-3}$ (corresponding to $\text{[M/H]}=-2.3$ and $-1.3$, respectively). 

More specifically, Fig.~\ref{fig:histo} shows the resulting number of stars $N$ with a given quality of fit, $\chi^2$, where for clarity we truncate the $\chi^2$ at 40. Overall, the most massive metal-poor AGB stars (5~M$_{\odot}$, and $Z=10^{-4}$) provide the best fit to the Sculptor abundances, which remains true if we expand the fitting range to all available abundance measurements of the $\alpha$- and iron peak elements (from \citealt{North12,Skuladottir15b,Skuladottir17,Hill18}). According to the MDF in Sculptor, the vast majority of AGB stars that contribute to the chemical enrichment of the ISM are expected to be in the range $-2.5<\text{[Fe/H]}<-1.5$ (e.g. \citealt{Battaglia08b}; Paper~I). This is consistent with the lower metallicity models in Fig.~\ref{fig:histo}, giving better match to the data. However, it is not immediately obvious why the higher mass AGB star models are favoured.

The comparison of theoretical yields to the abundance pattern of two typical Sculptor stars ($\text{[Fe/H]}=-1.9$, and $-0.9$) is shown in  Fig.~\ref{fig:twostars}. In addition to two AGB models from the F.R.U.I.T.Y database (2, and 5~M$_\odot$, with $Z=10^{-4}$), we also include a model for MHD jet SN \citep{Winteler12} and a NS-NS merger (each of 1~M$_{\odot}$ from \citealt{Korobkin12,Rosswog13}). Here, we only consider the dynamical ejecta, which is not a complete representation. However, for the elements in question a considerable contribution from the wind ejecta is not expected.

In the metal-poor star ET0048, at $\text{[Fe/H]}=-1.9$, the abundance pattern of the heavy $n$-capture elements (Ba, La, Nd, and Eu)  is consistent with mainly being the result of the $r$-process. Both MHD SN and NSM show reasonably good fits (see Fig.~\ref{fig:twostars}). This is consistent with the conclusion that AGB stars only start to become significant when the ISM in Sculptor has reached $\text{[Fe/H]}\approx-2$ (see Fig.~\ref{fig:ncap} and discussion in Paper~I). However, [Y/Ba] in ET0048 is higher than predicted by the $r$-process, suggesting a significant contribution from a LEPP (as shown in Fig.~\ref{fig:ncapba} and \ref{fig:logeps}). Another possibility is that the MHD SN is induced by the rotation of a massive star. In case of rotation, the lighter $n$-capture elements (e.g. Y) can be produced during its stellar lifetime (\citealt{Choplin18}; see also e.g. Fig.~\ref{fig:rrms}). A rotating massive star experiencing MHD SN may therefore be consistent with ET0048.

The metal-rich star ET0051, at $\text{[Fe/H]}=-0.9$ (bottom panel of Fig.~\ref{fig:twostars}), shows clearer signs of $s$-process influences, meaning lower [Eu/Ba], but also lower [La/Ba] and [Nd/Ba], which is typical for stars at higher metallicities in Sculptor (see Fig.~\ref{fig:ncapba}). 
Although we see that AGB yields provide better fits to ET0051 with respect to MHD SN and NS-NS, we cannot clearly identify a model that fully reproduce these abundances. While the [Y/Ba] ratio is more in line with low-mass AGB models, $M=2~$M$_\odot$, the [La,Nd/Ba] ratios are more compatible with the higher mass models, $M=5~$M$_\odot$. As this is a general property of the F.R.U.I.T.Y database, i.e. that higher mass stars have lower [La,Nd/Ba], and it results in the high-mass models giving better fits in Fig.~\ref{fig:histo}. Even in the 5~M$_\odot$ model, however, the predicted values of [La/Ba] are significantly higher than the observed values. This is discussed in more detail in Section~\ref{sec:si}.\\

              \begin{figure*}
   \centering
   \includegraphics[width=\hsize-2cm]{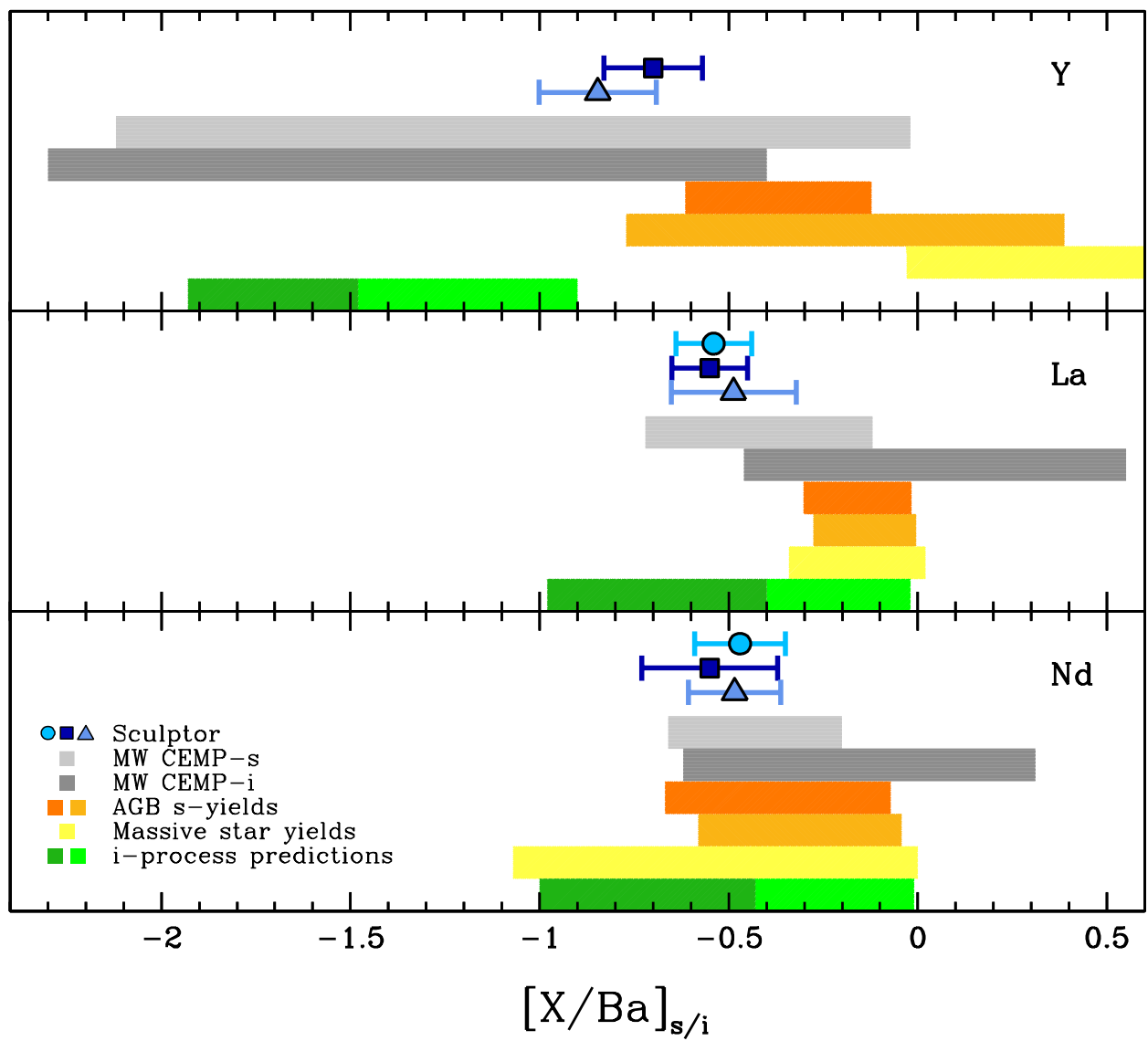}
      \caption{Blue symbols are the abundance ratios of the cumulative $s$- and/or $i$-process enrichment, i.e. [X/Ba]$_\text{s/i}$, in the chemical evolution of Sculptor from $\text{[Fe/H]}=-2$ to $-1$. Different symbols are three independent methods to remove the $r$-process from the measured abundances: 1)~Circles assume the $r$-process pattern from \citet{Bisterzo14}; 2)~Squares adopt the pattern of the strongly $r$-process enhanced star CS~22892-052 from \citet{Sneden03}; 3)~Triangles assume that any increase of [X/Mg] with [Fe/H] in Sculptor comes only from the $s$-process (see text for details). Errors include measurement errors in Sculptor and the errors from \citet{Bisterzo14}. The abundance range of Milky Way CEMP-$s$ stars with a undiluted $s$-process signature, $\text{[Eu/Ba]}<-0.9$, is shown with a light gray bar, and a dark gray bar is the same for CEMP-$i$ stars with $-0.5\leq\text{[Eu/Ba]}\leq0$, $\text{[Ba/Fe]}>1$, and $\text{[Eu/Fe]}>1$. The range of all available $s$-process models for AGB stars in the F.R.U.I.T.Y database are shown for two metallicities: $Z=10^{-4}$ (dark orange); and $Z=10^{-3}$ (light orange). Yellow is the range of abundance ratios for nonstandard $s$-process in massive stars, $10-150$~M$_\odot$ with $Z=10^{-3}$ \citep{Choplin18}. The range of the pure $i$-process at $Z=10^{-4}$ is in green, where light green refers to $n$-densities of $10^{12}\leq n_d/\text{cm}^{-3} \leq 10^{14}$, and dark green assumes higher $10^{14}<n_d/\text{cm}^{-3}\leq10^{15}$ \citep{Hampel16}. \textit{CEMP-$s$ references:} \citealt{Aoki02,Cohen13,Roederer14}; \textit{CEMP-$i$ references:} \citealt{Aoki02,Goswami06,Behara10,Cohen13,Cui13,HansenT15,Hampel19} (which includes the sample from \citealt{Abate15b}).
      }
         \label{fig:oy}
   \end{figure*}   

\subsection{Correcting for the r-process} \label{sec:rcorr}
   
The comparison of chemical abundance patterns in the Sculptor dSph with theoretical models is somewhat complicated by the fact that all processes, the $s$-, $i$- and $r$-processes, contribute to the overall build-up of $n$-capture elements. The predicted AGB stellar yields depend heavily on both the mass and metallicity of the star, and other model parameters such as rotation and the size of the $^{13}$C-pocket (e.g.~\citealt{Karakas14,Cristallo15}). The $r$-process, on the other hand, has been observationally proven to be quite robust for the elements from Ba to Eu (e.g. \citealt{Sneden03,Sneden08}). Therefore, we can estimate and subtract the $r$-process contribution from the build-up of $n$-capture elements in Sculptor.

To ensure that our results are robust, we explore three different methods to correct for the $r$-process, namely by assuming: 1)~the solar-scaled $r$-process pattern \citep{Bisterzo14}; 2)~the measured abundances of the $r$-process rich star CS~22892-052 \citep{Sneden03}; 3)~that the $r$-process and ccSN have similar timescales (see Paper~I), so that any increase in [X/Mg] (where X has $Z>30$) with [Fe/H] is due to a delayed $s$- and/or $i$-process.

These independent assumptions about the $r$-process give very consistent results (see Fig.~\ref{fig:oy}). One exception is [Y/Ba], while assuming the solar-scaled $r$-process \citep{Bisterzo14}. This predicts higher [Y/Ba] than observed in Sculptor, meaning that even if all Y came from the $r$-process, the predicted [Y/Ba]$_r$ ratios are not reached. It has been shown that there is a significant range of [Y/Ba] in metal-poor, pure-$r$ stars (e.g. \citealt{Roederer10}), which are typically much lower than those of the solar-scaled $r$-process (see also Fig.~\ref{fig:ncapba}). We therefore conclude that the [Y/Ba] ratios from \citet{Bisterzo14}, which are based on the solar abundance, are not appropriate for our sample. Otherwise, the overall good agreement between different methods, suggest that the measured [X/Ba]$_\textsl{s/i}$ in Sculptor are reliable.


\subsection{The signature of the $s$- and/or $i$-process enrichment} \label{sec:si}

By correcting for the $r$-process (Section~\ref{sec:rcorr}), we have isolated the abundance ratios of $\text{[X/Ba]}_\textsl{s/i}$, meaning the cumulative $s$- and/or $i$-process enrichment occurring from $\text{[Fe/H]}=-2$ to~$-1$ in the Sculptor dSph. Our results are shown in Fig.~\ref{fig:oy} along with the range of measured [X/Ba] in Milky Way pure CEMP-$s$ stars ($\text{[Eu/Ba]}<-0.9$), and CEMP-$i$ stars ($-0.5\leq\text{[Eu/Ba]}\leq0$). For comparison with theoretical models, we include the range of: a)~all available $s$-process AGB stellar yields in the F.R.U.I.T.Y database \citep{Cristallo09,Cristallo11}, for $Z=10^{-4}$ and $10^{-3}$, b)~yield predictions for massive stars ($10-150$~M$_\odot$) at $Z=10^{-3}$, with and without rotation \citep{Choplin18}, c)~predictions for pure $i$-processed material at $Z=10^{-4}$ \citep{Hampel16}.

From Fig.~\ref{fig:oy} (top panel) it is clear that the ratio of  [Y/Ba]$_{s/i}$ in Sculptor is at the higher end of [Y/Ba] in CEMP-$s$ and CEMP-$i$ stars in the Milky Way. This may in part be a consequence of the lower metallicity of the CEMP sample, $-3<{\text{[Fe/H]}}_\text{CEMP}<-2$, relative to Sculptor, which has an MDF that peaks around $\text{[Fe/H]}\approx-2$, so the majority of the stars have $-2.5<\text{[Fe/H]}<-1.5$ (e.g.~\citealt{Battaglia08b}; Paper~I). The AGB stellar yields for [Y/Ba] are predicted to be quite metallicity-dependent, i.e. increasing towards higher metallicities (e.g.~\citealt{Karakas14,Cristallo15}). However, we note that there is no statistically significant increase of [Y/Ba] with [Fe/H], within our selected sample of CEMP-$s$ stars. Another difference between these samples is the fact that the CEMP-$s$ and -$i$ stars trace enrichment by individual stars in binary systems while the Sculptor values should be interpreted as integrated yield values over the stellar IMF. In other words, these samples include different mass distributions of contributing AGB stars, which can cause differences in the resulting [X/Ba], as opposed to the yields of just one single AGB star which are observed in CEMP-$s$ and -$i$ stars.

The observed value of [Y/Ba]$_{s/i}$ in Sculptor is lower than expected in relation to the predicted $s$-process AGB stellar yields from the F.R.U.I.T.Y database, see Fig.~\ref{fig:oy}. Furthermore, the disagreement of the yields with Milky Way CEMP-$s$ and -$i$ stars is even stronger. We note that this result is not limited to our choice of $s$-process yields. \citet{Lugaro12} predict $\text{[Sr/Ba]}\gtrsim-1$, and were thus unable to reproduce the lowest [Sr/Ba] in their comparison samples of observed CEMP-$s$, and -$i$ stars. Since the models of massive stars predict even larger [Y/Ba], it is unlikely that these are the dominant source of $s$- and $i$-process elements in the cumulative enrichment of the Sculptor dSph. However, results of the pure $i$-process predict much lower [Y/Ba], consistent with the bulk of CEMP-$s$, and -$i$ stars \citep{Hampel16}. The [Y/Ba]$_{s/i}$ ratio in Sculptor is thus best explained with a combination of the $s$- and $i$-processes.

The ratio of the neighboring elements, [La/Ba]$_{s/i}$, in Sculptor is consistent with CEMP-$s$ stars, but lower than those of CEMP-$i$ stars, see Fig.~\ref{fig:oy}. Similarly to Y, the $s$-process yields for [La/Ba] are higher than the measured value in Sculptor, both for AGB and massive stars. Furthermore, they are not able to reproduce the full range of [La/Ba] ratios in the pure CEMP-$s$ stars. The theoretical $i$-process is, however, consistent with the low values of Sculptor and CEMP-$s$ stars. We note, that at models at the highest $n$-density, $n_d=10^{15}$\,cm$^{-3}$, predict $\text{[Eu/Ba]}=-1$, which would observationally be defined as the $s$-process. Thus, some of the stars in our CEMP-$s$ sample, might actually be CEMP-$i$ stars. Curiously, the pure $i$-process models are not able to explain the high $\text{[La/Ba]}>0$, which are observed in some Milky Way CEMP-$i$ stars (e.g.~\citealt{Goswami06,Cohen13,HansenT15}). One possible explanation is that these could actually be CEMP-$r/s$ stars, meaning that they show a combined pollution by both the $r$- and $s$-processes, since the $r$-process has $\text{[La/Ba]}>0$ (see Fig.~\ref{fig:ncapba}).

All $s$- and $i$-process models in Fig.~\ref{fig:oy} agree with the measured value of [Nd/Ba]$_{s/i}$ in the Sculptor dSph, as well as the range of CEMP-$s$ stars in the Milky Way. However, like in the case of La, the pure $i$-process models are not consistent with the high [Nd/Ba] in the Milky Way CEMP-$i$ stars (e.g.~\citealt{Goswami06,Hampel19}).

Finally, we note that since the $i$-process can reach very low values, $\text{[Eu/Ba]}\approx-1$ \citep{Hampel16}, the [Eu/Ba] ratio (as suggested by \citealt{Beers05}) is probably not a good indicator to separate between the $s$- and $i$-process, since both processes overlap significantly in this space  \citep{Abate15c}. Other probes, such as a combination of [La/Eu] and [Hf/Ir], have been proposed \citep{Frebel18}, but since both Hf and Ir are very challenging to measure, this definition is not very practical. However, looking at the top panel of Fig.~\ref{fig:oy}, the theoretical predictions of the $s$- and $i$-processes separate nicely in [Y/Ba], or equivalently in [Sr/Ba], which \citet{Hansen19} suggested as a very useful indicator of different CEMP subclasses.

Overall, the abundance pattern in the Sculptor dSph (Fig.~\ref{fig:oy}) cannot be explained by the $s$-process alone, but is better reproduced by a combination of the $s$- and $i$-processes. However, the abundance ratios in Fig.~\ref{fig:oy} assume that the $r$-process is the dominant production site of Eu in Sculptor, which is supported by the flat trend of 
$\text{[Eu/Mg]}\approx0$ at $-2.4\leq\text{[Fe/H]}\leq-0.9$ in Sculptor (Paper~I). However, if a significant fraction of Eu is created by the $i$-process, this would cause an increase in [Eu/Mg] with [Fe/H]. To quantitatively constrain how much the $i$-process contributes to the build-up of $n$-capture elements in Sculptor, a full set of $i$-process yields for a range of AGB stellar masses is necessary. Unfortunately, this is unavailable at present. In addition, super-AGB stars have proton ingestion episodes, at all metalicities, but there is also no grid of those yet \citep{Doherty15,Jones16}. A full chemical evolution modeling is therefore not possible at this point. However, our current results confirm that the $i$-process is an important source in the overall evolution of the heavy elements in the Sculptor dSph.

\section{Discussion \& Conclusions}

The detailed chemical abundances of Mg, Y, Ba, La, Nd, and Eu in 98 stars in Sculptor have allowed us to explore the relative contribution of the different $n$-capture processes in this dSph galaxy: the slow ($s$), intermediate ($i$), and rapid ($r$) processes. Our results show that while the $r$-process is consistent with having little or no time delay relative to ccSN, the characteristic timescale of AGB stars is comparable with that of SN type~Ia. 

The influences of AGB stars in Sculptor become significant at $\text{[Fe/H]}\approx-2$. The chemical imprint of AGB stars causes [Y/Ba], [La/Ba], [Nd/Ba], and [Eu/Ba] to decrease with metallicity, reaching subsolar values, $\text{[X/Ba]}<0$, at the highest $\text{[Fe/H]}=-1$ (Section~\ref{sec:ba}). At $\text{[Fe/H]}\gtrsim-1.4$, the contribution of AGB stars, relative to the $r$-process, is stronger in Sculptor than in the Milky Way. This is likely caused by the fact that star formation in Sculptor progressed slowly, and decreased until it was completely quenched when the galaxy lost all its gas. At later times the chemical enrichment was thus dominated by delayed processes, most notably AGB stars and SN type Ia. Presumably, the products of these delayed nucleosynthetic sites were released into a small mass of gas, further enhancing their effects (see also \citealt{Salvadori15}).

There is a break in the trend of [Y/H] vs [Ba/H] when AGB stars start to dominate the production of both elements (Section~\ref{sec:yba}). At low metallicities, $\text{[Fe/H]}<-2$, [Y/Ba] follows the upper envelope of the same ratios in the Milky Way. This has proven to be typical for dSph galaxies, while ultra-faint dwarf galaxies typically have lower values \citep{Mashonkina17}. Similarly to the Milky Way, it is not clear what process causes the supersolar $\text{[Y/Ba]}>0$ at the lowest $\text{[Ba/H]}<-2.8$, but possible explanations include massive stars with rotation, or the LEPP/weak-$r$/limited-$r$ process.

The ISM in Sculptor seems to have been well mixed at $\text{[Fe/H]}\gtrsim-2$, judging by the relatively small scatter in the chemical abundances. However, correlations in the scatter of different $n$-capture elements suggest that some internal scatter in these element might be present at any given [Fe/H], i.e. that the heavy elements were well but not completely homogeneously mixed into the ISM at the center of Sculptor (Section~\ref{sec:scatter}). This applies both to a) the products of AGB stars, which are released by stellar winds which are much less energetic than SNe, and are thus more likely to be inhomogeneously mixed; and b) the $r$-process, where the rarity of the events is a likely cause of any ISM inhomogeneities.

To shed light on possible individual sources of heavy elements in Sculptor, we looked at three stars from the literature, which are rich in $n$-elements, and compared them to theoretical yield predictions of single NSM, MHD SN, AGB, and massive stars (including rotation). For one of these stars (S-400), no satisfactory fit was found (Section~\ref{sec:pec}), and its enhanced abundance pattern of the $n$-capture elements might therefore be the result of more than one $n$-capture process.  The abundance pattern of the likely CH star G-982 at $\text{[Fe/H]}=-1$ from \citet{Geisler05} is very well reproduced with a model of the $i$-process, though very high $\text{[Pb/Fe]}\approx4$ are predicted. 

Finally, the CEMP-no star ET0097 at $\text{[Fe/H]}=-2$, which has unusually high $\text{[Sr,Y,Zr/Ba]}>+0.7$ \citep{Skuladottir15a}, was best fit by a rapidly rotating massive star, which was able to explain both the high [C/Fe] and the high lighter $n$-capture elements. Similar conclusions were reached by \citet{Yong17} for the star ROA~276, in $\omega$~Centauri, which has extremely high $\text{[X/Fe]}>+1$ of the lighter $n$-capture elements ($29<Z<43$). However the massive star models were not able to reproduce the low $\text{[C/Fe]}<0$ in ROA~276. 
The apparently higher frequencies of CEMP-no stars with enhanced $\text{[Sr,Y,Zr/Ba}]>+0.7$ in dwarf galaxies compared to the Milky Way (see Section~\ref{sec:intro}), might thus indicate that the products of massive stars are more commonly released into a small body of gas in these small systems compared to the Milky Way, causing the very clear imprint of their yields.

We measured the cumulative build-up of the $n$-capture elements in Sculptor, and isolated the $s$-, and/or $i$-process component. The correction for the $r$-process was done with three independent methods, which all gave consistent results (see Section~\ref{sec:rcorr}). Thus, for the first time, we were able to measure the chemical abundance ratios in the cumulative $s$-, and/or $i$-process in Sculptor: $\text{[Y/Ba]}_{s/i}=-0.85\pm0.16$, $\text{[La/Ba]}_{s/i}=-0.49\pm0.17$, and $\text{[Nd/Ba]}_{s/i}=-0.48\pm0.12$. This is an important constraint of IMF integrated yields at low [Fe/H], and thus highly complementary to studies of CEMP-$s$, and -$i$ stars that experienced mass transfer from an individual AGB binary companion. 

Overall, our measured [X/Ba]$_{s/i}$ agree well with those of CEMP-$s$, and -$i$ stars in the Milky Way. However, when comparing to theoretical yields of the $s$-process in AGB or massive stars, it is clear that they cannot reproduce the low $\text{[Y/Ba]}_{s/i}$ and $\text{[La/Ba]}_{s/i}$ measured in Sculptor (Section~\ref{sec:si}). This tension can be alleviated by also taking into account the contribution of the $i$-process, which predicts much lower [Y/Ba] and [La/Ba] compared to the $s$-process. Thus, we conclude that \textit{both the $s$- and the $i$-process were important sources for the chemical enrichment in the Sculptor dSph galaxy}.

However, if the $i$-process in AGB stars also created significant amounts of Eu in Sculptor, an increase of [Eu/Mg] with [Fe/H] would be expected, which is not observed in the currently available data (see Paper~I). To fully quantify how much the $i$-process contributed to the overall build-up of the $n$-capture elements in Sculptor, a complete set of $i$-process yields are needed, both for super AGB stars \citep{Doherty15,Doherty17,Jones16} as well as low-mass metal-poor stars \citep{Campbell10,Cristallo16}.

We note that the theoretical predictions for the pure $i$-process are currently not able to reproduce the highest $\text{[La/Ba]}>0$, and $\text{[Nd/Ba]}>0$, which are observed in a subsample of Milky Way CEMP-$i$ stars. These might, therefore, actually be CEMP-$r/s$ stars, showing a combined signature of the $s$- and $r$-processes. Finally, given that models of the $i$-process predict quite low $\text{[Eu/Ba]}\approx-1$ at high $n$-densities, $n_d=10^{15}$\,cm$^{-3}$, this abundance ratio might not be optimal to distinguish between CEMP-$s$, and -$i$ stars. Based on theoretical yield predictions, [Sr,Y,Zr/Ba] might be a better option, as was proposed by \citet{Hansen19}. 

Ultimately, our new measurements of Y, Ba, La, Nd, and Eu in RGB stars in the Sculptor dSph galaxy, and comparison to theory, strongly suggest that the $i$-process is important at low metallicity, and highlight that our understanding of the $n$-capture processes in the early universe is far from complete. \\

\begin{acknowledgements}
\'A.S.~acknowledges funds from the Alexander von Humboldt Foundation in the framework of the Sofja Kovalevskaja Award endowed by the Federal Ministry of Education and Research. This project has received funding from the European Research Council (ERC) under the European Union's Horizon 2020 research and innovation programme (grant agreement No. 804240). C.J.H acknowledges support from Max Planck. This article is based upon work from the ``ChETEC" COST Action (CA16117), supported by COST (European Cooperation in Science and Technology). A.C.~acknowledges funding from the Swiss National Science Foundation under grant P2GEP2\_184492. S.W.C acknowledges federal funding from the Australian Research Council though a Future Fellowship (FT160100046), and Discovery Project (DP190102431). This work was supported in part by resources provided by the Governments of Australia and Western Australia, via the National Computational Merit Allocation Scheme (projects ew6 and ke38). We also thank E.~Tolstoy for insightful suggestions. 
\end{acknowledgements}

\bibliography{heimildir}

\begin{thebibliography}{150}
\expandafter\ifx\csname natexlab\endcsname\relax\def\natexlab#1{#1}\fi

\bibitem[{{Abate} {et~al.}(2015{\natexlab{a}}){Abate}, {Pols}, {Izzard}, \&
  {Karakas}}]{Abate15b}
{Abate}, C., {Pols}, O.~R., {Izzard}, R.~G., \& {Karakas}, A.~I.
  2015{\natexlab{a}}, \aap, 581, A22

\bibitem[{{Abate} {et~al.}(2015{\natexlab{b}}){Abate}, {Pols}, {Karakas}, \&
  {Izzard}}]{Abate15a}
{Abate}, C., {Pols}, O.~R., {Karakas}, A.~I., \& {Izzard}, R.~G.
  2015{\natexlab{b}}, \aap, 576, A118

\bibitem[{{Abate} {et~al.}(2015{\natexlab{c}}){Abate}, {Pols}, {Karakas}, \&
  {Izzard}}]{Abate15c}
{Abate}, C., {Pols}, O.~R., {Karakas}, A.~I., \& {Izzard}, R.~G.
  2015{\natexlab{c}}, \aap, 576, A118

\bibitem[{{Abate} {et~al.}(2018){Abate}, {Pols}, \& {Stancliffe}}]{Abate18}
{Abate}, C., {Pols}, O.~R., \& {Stancliffe}, R.~J. 2018, \aap, 620, A63

\bibitem[{{Allen} {et~al.}(2012){Allen}, {Ryan}, {Rossi}, {Beers}, \&
  {Tsangarides}}]{Allen12}
{Allen}, D.~M., {Ryan}, S.~G., {Rossi}, S., {Beers}, T.~C., \& {Tsangarides},
  S.~A. 2012, \aap, 548, A34

\bibitem[{{Amarsi} {et~al.}(2019{\natexlab{a}}){Amarsi}, {Nissen}, {Asplund},
  {Lind}, \& {Barklem}}]{Amarsi19}
{Amarsi}, A.~M., {Nissen}, P.~E., {Asplund}, M., {Lind}, K., \& {Barklem},
  P.~S. 2019{\natexlab{a}}, \aap, 622, L4

\bibitem[{{Amarsi} {et~al.}(2019{\natexlab{b}}){Amarsi}, {Nissen}, \&
  {Sk{\'u}lad{\'o}ttir}}]{AmarsiNS19}
{Amarsi}, A.~M., {Nissen}, P.~E., \& {Sk{\'u}lad{\'o}ttir}, {\'A}.
  2019{\natexlab{b}}, \aap, 630, A104

\bibitem[{{Andrievsky} {et~al.}(2017){Andrievsky}, {Korotin}, {Hill}, \&
  {Zhukova}}]{Andrievsky17}
{Andrievsky}, S.~M., {Korotin}, S.~A., {Hill}, V., \& {Zhukova}, A.~V. 2017,
  arXiv e-prints

\bibitem[{{Andrievsky} {et~al.}(2011){Andrievsky}, {Spite}, {Korotin},
  {Fran{\c{c}}ois}, {Spite}, {Bonifacio}, {Cayrel}, \& {Hill}}]{Andrievsky11}
{Andrievsky}, S.~M., {Spite}, F., {Korotin}, S.~A., {et~al.} 2011, \aap, 530,
  A105

\bibitem[{{Andrievsky} {et~al.}(2010){Andrievsky}, {Spite}, {Korotin}, {Spite},
  {Bonifacio}, {Cayrel}, {Fran{\c{c}}ois}, \& {Hill}}]{Andrievsky10}
{Andrievsky}, S.~M., {Spite}, M., {Korotin}, S.~A., {et~al.} 2010, \aap, 509,
  A88

\bibitem[{{Andrievsky} {et~al.}(2007){Andrievsky}, {Spite}, {Korotin}, {Spite},
  {Bonifacio}, {Cayrel}, {Hill}, \& {Fran{\c{c}}ois}}]{Andrievsky07}
{Andrievsky}, S.~M., {Spite}, M., {Korotin}, S.~A., {et~al.} 2007, \aap, 464,
  1081

\bibitem[{{Andrievsky} {et~al.}(2009){Andrievsky}, {Spite}, {Korotin}, {Spite},
  {Fran{\c{c}}ois}, {Bonifacio}, {Cayrel}, \& {Hill}}]{Andrievsky09}
{Andrievsky}, S.~M., {Spite}, M., {Korotin}, S.~A., {et~al.} 2009, \aap, 494,
  1083

\bibitem[{{Aoki} {et~al.}(2007){Aoki}, {Beers}, {Christlieb}, {Norris}, {Ryan},
  \& {Tsangarides}}]{Aoki07}
{Aoki}, W., {Beers}, T.~C., {Christlieb}, N., {et~al.} 2007, \apj, 655, 492

\bibitem[{{Aoki} {et~al.}(2002){Aoki}, {Ryan}, {Norris}, {Beers}, {Ando}, \&
  {Tsangarides}}]{Aoki02}
{Aoki}, W., {Ryan}, S.~G., {Norris}, J.~E., {et~al.} 2002, \apj, 580, 1149

\bibitem[{{Arcones} \& {Montes}(2011)}]{Arcones11}
{Arcones}, A. \& {Montes}, F. 2011, \apj, 731, 5

\bibitem[{{Arentsen} {et~al.}(2019){Arentsen}, {Starkenburg}, {Shetrone},
  {Venn}, {Depagne}, \& {McConnachie}}]{Arentsen19}
{Arentsen}, A., {Starkenburg}, E., {Shetrone}, M.~D., {et~al.} 2019, \aap, 621,
  A108

\bibitem[{{Banerjee} {et~al.}(2018){Banerjee}, {Qian}, \& {Heger}}]{Banerjee18}
{Banerjee}, P., {Qian}, Y.-Z., \& {Heger}, A. 2018, \apj, 865, 120

\bibitem[{{Battaglia} {et~al.}(2008{\natexlab{a}}){Battaglia}, {Helmi},
  {Tolstoy}, {Irwin}, {Hill}, \& {Jablonka}}]{Battaglia08a}
{Battaglia}, G., {Helmi}, A., {Tolstoy}, E., {et~al.} 2008{\natexlab{a}},
  \apjl, 681, L13

\bibitem[{{Battaglia} {et~al.}(2008{\natexlab{b}}){Battaglia}, {Irwin},
  {Tolstoy}, {Hill}, {Helmi}, {Letarte}, \& {Jablonka}}]{Battaglia08b}
{Battaglia}, G., {Irwin}, M., {Tolstoy}, E., {et~al.} 2008{\natexlab{b}},
  \mnras, 383, 183

\bibitem[{{Beers} \& {Christlieb}(2005)}]{Beers05}
{Beers}, T.~C. \& {Christlieb}, N. 2005, \araa, 43, 531

\bibitem[{{Behara} {et~al.}(2010){Behara}, {Bonifacio}, {Ludwig}, {Sbordone},
  {Gonz{\'a}lez Hern{\'a}ndez}, \& {Caffau}}]{Behara10}
{Behara}, N.~T., {Bonifacio}, P., {Ludwig}, H.-G., {et~al.} 2010, \aap, 513,
  A72

\bibitem[{{Bergemann} {et~al.}(2012){Bergemann}, {Hansen}, {Bautista}, \&
  {Ruchti}}]{Bergemann12}
{Bergemann}, M., {Hansen}, C.~J., {Bautista}, M., \& {Ruchti}, G. 2012, \aap,
  546, A90

\bibitem[{{Bergemann} {et~al.}(2015){Bergemann}, {Kudritzki}, {Gazak},
  {Davies}, \& {Plez}}]{Bergemann15}
{Bergemann}, M., {Kudritzki}, R.-P., {Gazak}, Z., {Davies}, B., \& {Plez}, B.
  2015, \apj, 804, 113

\bibitem[{{Bettinelli} {et~al.}(2019){Bettinelli}, {Hidalgo}, {Cassisi},
  {Aparicio}, {Piotto}, {Valdes}, \& {Walker}}]{Bettinelli19}
{Bettinelli}, M., {Hidalgo}, S.~L., {Cassisi}, S., {et~al.} 2019, \mnras, 1642

\bibitem[{{Bisterzo} {et~al.}(2014){Bisterzo}, {Travaglio}, {Gallino},
  {Wiescher}, \& {K{\"a}ppeler}}]{Bisterzo14}
{Bisterzo}, S., {Travaglio}, C., {Gallino}, R., {Wiescher}, M., \&
  {K{\"a}ppeler}, F. 2014, \apj, 787, 10

\bibitem[{{Burbidge} {et~al.}(1957){Burbidge}, {Burbidge}, {Fowler}, \&
  {Hoyle}}]{Burbidge57}
{Burbidge}, E.~M., {Burbidge}, G.~R., {Fowler}, W.~A., \& {Hoyle}, F. 1957,
  Reviews of Modern Physics, 29, 547

\bibitem[{Burris {et~al.}(2000)Burris, Pilachowski, Armandroff, Sneden, Cowan,
  \& Roe}]{Burris00}
Burris, D., Pilachowski, C., Armandroff, T., {et~al.} 2000, ApJ, 544, 302

\bibitem[{{Cameron}(1957)}]{Cameron57}
{Cameron}, A.~G.~W. 1957, \pasp, 69, 201

\bibitem[{{Campbell} \& {Lattanzio}(2008)}]{Campbell08}
{Campbell}, S.~W. \& {Lattanzio}, J.~C. 2008, \aap, 490, 769

\bibitem[{{Campbell} {et~al.}(2010){Campbell}, {Lugaro}, \&
  {Karakas}}]{Campbell10}
{Campbell}, S.~W., {Lugaro}, M., \& {Karakas}, A.~I. 2010, \aap, 522, L6

\bibitem[{{Chiappini} {et~al.}(2011){Chiappini}, {Frischknecht}, {Meynet},
  {Hirschi}, {Barbuy}, {Pignatari}, {Decressin}, \& {Maeder}}]{Chiappini11}
{Chiappini}, C., {Frischknecht}, U., {Meynet}, G., {et~al.} 2011, \nat, 472,
  454

\bibitem[{{Choplin} {et~al.}(2017){Choplin}, {Hirschi}, {Meynet}, \&
  {Ekstr{\"o}m}}]{Choplin17}
{Choplin}, A., {Hirschi}, R., {Meynet}, G., \& {Ekstr{\"o}m}, S. 2017, \aap,
  607, L3

\bibitem[{{Choplin} {et~al.}(2018){Choplin}, {Hirschi}, {Meynet},
  {Ekstr{\"o}m}, {Chiappini}, \& {Laird}}]{Choplin18}
{Choplin}, A., {Hirschi}, R., {Meynet}, G., {et~al.} 2018, \aap, 618, A133

\bibitem[{{Clarkson} {et~al.}(2018){Clarkson}, {Herwig}, \&
  {Pignatari}}]{Clarkson18}
{Clarkson}, O., {Herwig}, F., \& {Pignatari}, M. 2018, \mnras, 474, L37

\bibitem[{{Cohen} {et~al.}(2013){Cohen}, {Christlieb}, {Thompson}, {McWilliam},
  {Shectman}, {Reimers}, {Wisotzki}, \& {Kirby}}]{Cohen13}
{Cohen}, J.~G., {Christlieb}, N., {Thompson}, I., {et~al.} 2013, \apj, 778, 56

\bibitem[{{C{\^o}t{\'e}} {et~al.}(2019){C{\^o}t{\'e}}, {Eichler}, {Arcones},
  {Hansen}, {Simonetti}, {Frebel}, {Fryer}, {Pignatari}, {Reichert},
  {Belczynski}, \& {Matteucci}}]{Cote19}
{C{\^o}t{\'e}}, B., {Eichler}, M., {Arcones}, A., {et~al.} 2019, \apj, 875, 106

\bibitem[{{Cowan} \& {Rose}(1977)}]{CowanRose77}
{Cowan}, J.~J. \& {Rose}, W.~K. 1977, \apj, 212, 149

\bibitem[{{Cristallo} {et~al.}(2016){Cristallo}, {Karinkuzhi}, {Goswami},
  {Piersanti}, \& {Gobrecht}}]{Cristallo16}
{Cristallo}, S., {Karinkuzhi}, D., {Goswami}, A., {Piersanti}, L., \&
  {Gobrecht}, D. 2016, \apj, 833, 181

\bibitem[{{Cristallo} {et~al.}(2011){Cristallo}, {Piersanti}, {Straniero},
  {Gallino}, {Dom{\'{\i}}nguez}, {Abia}, {Di Rico}, {Quintini}, \&
  {Bisterzo}}]{Cristallo11}
{Cristallo}, S., {Piersanti}, L., {Straniero}, O., {et~al.} 2011, \apjs, 197,
  17

\bibitem[{{Cristallo} {et~al.}(2009){Cristallo}, {Straniero}, {Gallino},
  {Piersanti}, {Dom{\'{\i}}nguez}, \& {Lederer}}]{Cristallo09}
{Cristallo}, S., {Straniero}, O., {Gallino}, R., {et~al.} 2009, \apj, 696, 797

\bibitem[{{Cristallo} {et~al.}(2015){Cristallo}, {Straniero}, {Piersanti}, \&
  {Gobrecht}}]{Cristallo15}
{Cristallo}, S., {Straniero}, O., {Piersanti}, L., \& {Gobrecht}, D. 2015,
  \apjs, 219, 40

\bibitem[{{Cruz} {et~al.}(2013){Cruz}, {Serenelli}, \& {Weiss}}]{Cruz13}
{Cruz}, M.~A., {Serenelli}, A., \& {Weiss}, A. 2013, \aap, 559, A4

\bibitem[{{Cui} {et~al.}(2013){Cui}, {Sivarani}, \& {Christlieb}}]{Cui13}
{Cui}, W.~Y., {Sivarani}, T., \& {Christlieb}, N. 2013, \aap, 558, A36

\bibitem[{{Dardelet} {et~al.}(2015){Dardelet}, {Ritter}, {Prado}, {Heringer},
  {Higgs}, {Sandalski}, {Jones}, {Denissenkov}, {Venn}, {Bertolli},
  {Pignatari}, {Woodward}, \& {Herwig}}]{Dardelet15}
{Dardelet}, L., {Ritter}, C., {Prado}, P., {et~al.} 2015, arXiv e-prints,
  arXiv:1505.05500

\bibitem[{{de Bennassuti} {et~al.}(2017){de Bennassuti}, {Salvadori},
  {Schneider}, {Valiante}, \& {Omukai}}]{deBennassuti17}
{de Bennassuti}, M., {Salvadori}, S., {Schneider}, R., {Valiante}, R., \&
  {Omukai}, K. 2017, \mnras, 465, 926

\bibitem[{{de Boer} {et~al.}(2012){de Boer}, {Tolstoy}, {Hill}, {Saha},
  {Olsen}, {Starkenburg}, {Lemasle}, {Irwin}, \& {Battaglia}}]{deBoer12}
{de Boer}, T.~J.~L., {Tolstoy}, E., {Hill}, V., {et~al.} 2012, \aap, 539, A103

\bibitem[{{Denissenkov} {et~al.}(2017){Denissenkov}, {Herwig}, {Battino},
  {Ritter}, {Pignatari}, {Jones}, \& {Paxton}}]{Denissenkov17}
{Denissenkov}, P.~A., {Herwig}, F., {Battino}, U., {et~al.} 2017, \apjl, 834,
  L10

\bibitem[{{Doherty} {et~al.}(2017){Doherty}, {Gil-Pons}, {Siess}, \&
  {Lattanzio}}]{Doherty17}
{Doherty}, C.~L., {Gil-Pons}, P., {Siess}, L., \& {Lattanzio}, J.~C. 2017,
  \pasa, 34, e056

\bibitem[{{Doherty} {et~al.}(2015){Doherty}, {Gil-Pons}, {Siess}, {Lattanzio},
  \& {Lau}}]{Doherty15}
{Doherty}, C.~L., {Gil-Pons}, P., {Siess}, L., {Lattanzio}, J.~C., \& {Lau}, H.
  H.~B. 2015, \mnras, 446, 2599

\bibitem[{{Fishlock} {et~al.}(2014){Fishlock}, {Karakas}, {Lugaro}, \&
  {Yong}}]{Fishlock14}
{Fishlock}, C.~K., {Karakas}, A.~I., {Lugaro}, M., \& {Yong}, D. 2014, \apj,
  797, 44

\bibitem[{{Fran{\c c}ois} {et~al.}(2007){Fran{\c c}ois}, {Depagne}, {Hill},
  {Spite}, {Spite}, {Plez}, {Beers}, {Andersen}, {James}, {Barbuy}, {Cayrel},
  {Bonifacio}, {Molaro}, {Nordstr{\"o}m}, \& {Primas}}]{Francois07}
{Fran{\c c}ois}, P., {Depagne}, E., {Hill}, V., {et~al.} 2007, \aap, 476, 935

\bibitem[{{Frebel}(2018)}]{Frebel18}
{Frebel}, A. 2018, Annual Review of Nuclear and Particle Science, 68, 237

\bibitem[{{Freiburghaus} {et~al.}(1999){Freiburghaus}, {Rosswog}, \&
  {Thielemann}}]{Freiburghaus99}
{Freiburghaus}, C., {Rosswog}, S., \& {Thielemann}, F.~K. 1999, \apjl, 525,
  L121

\bibitem[{{Frischknecht} {et~al.}(2016){Frischknecht}, {Hirschi}, {Pignatari},
  {Maeder}, {Meynet}, {Chiappini}, {Thielemann}, {Rauscher}, {Georgy}, \&
  {Ekstr{\"o}m}}]{Frischknecht16}
{Frischknecht}, U., {Hirschi}, R., {Pignatari}, M., {et~al.} 2016, \mnras, 456,
  1803

\bibitem[{{Frischknecht} {et~al.}(2012){Frischknecht}, {Hirschi}, \&
  {Thielemann}}]{Frischknecht12}
{Frischknecht}, U., {Hirschi}, R., \& {Thielemann}, F.~K. 2012, \aap, 538, L2

\bibitem[{{Geisler} {et~al.}(2005){Geisler}, {Smith}, {Wallerstein},
  {Gonzalez}, \& {Charbonnel}}]{Geisler05}
{Geisler}, D., {Smith}, V.~V., {Wallerstein}, G., {Gonzalez}, G., \&
  {Charbonnel}, C. 2005, \aj, 129, 1428

\bibitem[{{Goswami} {et~al.}(2006){Goswami}, {Aoki}, {Beers}, {Christlieb},
  {Norris}, {Ryan}, \& {Tsangarides}}]{Goswami06}
{Goswami}, A., {Aoki}, W., {Beers}, T.~C., {et~al.} 2006, \mnras, 372, 343

\bibitem[{{Gratton} {et~al.}(2000){Gratton}, {Sneden}, {Carretta}, \&
  {Bragaglia}}]{Gratton00}
{Gratton}, R.~G., {Sneden}, C., {Carretta}, E., \& {Bragaglia}, A. 2000, \aap,
  354, 169

\bibitem[{{Hampel} {et~al.}(2019){Hampel}, {Karakas}, {Stancliffe}, {Meyer}, \&
  {Lugaro}}]{Hampel19}
{Hampel}, M., {Karakas}, A.~I., {Stancliffe}, R.~J., {Meyer}, B.~S., \&
  {Lugaro}, M. 2019, arXiv e-prints, arXiv:1910.11882

\bibitem[{{Hampel} {et~al.}(2016){Hampel}, {Stancliffe}, {Lugaro}, \&
  {Meyer}}]{Hampel16}
{Hampel}, M., {Stancliffe}, R.~J., {Lugaro}, M., \& {Meyer}, B.~S. 2016, \apj,
  831, 171

\bibitem[{{Hansen} {et~al.}(2019){Hansen}, {Hansen}, {Koch}, {Beers},
  {Nordstr{\"o}m}, {Placco}, \& {Andersen}}]{Hansen19}
{Hansen}, C.~J., {Hansen}, T.~T., {Koch}, A., {et~al.} 2019, \aap, 623, A128

\bibitem[{{Hansen} {et~al.}(2014){Hansen}, {Montes}, \& {Arcones}}]{Hansen14}
{Hansen}, C.~J., {Montes}, F., \& {Arcones}, A. 2014, \apj, 797, 123

\bibitem[{{Hansen} {et~al.}(2012){Hansen}, {Primas}, {Hartman}, {Kratz},
  {Wanajo}, {Leibundgut}, {Farouqi}, {Hallmann}, {Christlieb}, \&
  {Nilsson}}]{Hansen12}
{Hansen}, C.~J., {Primas}, F., {Hartman}, H., {et~al.} 2012, \aap, 545, A31

\bibitem[{{Hansen} {et~al.}(2015){Hansen}, {Hansen}, {Christlieb}, {Beers},
  {Yong}, {Bessell}, {Frebel}, {Garc{\'{\i}}a P{\'e}rez}, {Placco}, {Norris},
  \& {Asplund}}]{HansenT15}
{Hansen}, T., {Hansen}, C.~J., {Christlieb}, N., {et~al.} 2015, \apj, 807, 173

\bibitem[{{Hansen} {et~al.}(2016{\natexlab{a}}){Hansen}, {Andersen},
  {Nordstr{\"o}m}, {Beers}, {Placco}, {Yoon}, \& {Buchhave}}]{HansenT16no}
{Hansen}, T.~T., {Andersen}, J., {Nordstr{\"o}m}, B., {et~al.}
  2016{\natexlab{a}}, \aap, 586, A160

\bibitem[{{Hansen} {et~al.}(2016{\natexlab{b}}){Hansen}, {Andersen},
  {Nordstr{\"o}m}, {Beers}, {Placco}, {Yoon}, \& {Buchhave}}]{HansenT16s}
{Hansen}, T.~T., {Andersen}, J., {Nordstr{\"o}m}, B., {et~al.}
  2016{\natexlab{b}}, \aap, 588, A3

\bibitem[{{Herwig} {et~al.}(2011){Herwig}, {Pignatari}, {Woodward}, {Porter},
  {Rockefeller}, {Fryer}, {Bennett}, \& {Hirschi}}]{Herwig11}
{Herwig}, F., {Pignatari}, M., {Woodward}, P.~R., {et~al.} 2011, \apj, 727, 89

\bibitem[{{Herwig} {et~al.}(2014){Herwig}, {Woodward}, {Lin}, {Knox}, \&
  {Fryer}}]{Herwig14}
{Herwig}, F., {Woodward}, P.~R., {Lin}, P.-H., {Knox}, M., \& {Fryer}, C. 2014,
  \apjl, 792, L3

\bibitem[{{Hill} {et~al.}(2019){Hill}, {Sk{\'u}lad{\'o}ttir}, {Tolstoy},
  {Venn}, {Shetrone}, {Jablonka}, {Primas}, {Battaglia}, {de Boer}, \&
  {Fran{\c{c}}ois}}]{Hill18}
{Hill}, V., {Sk{\'u}lad{\'o}ttir}, {\'A}., {Tolstoy}, E., {et~al.} 2019, \aap,
  626, A15

\bibitem[{{Hirai} {et~al.}(2019){Hirai}, {Wanajo}, \& {Saitoh}}]{Hirai19}
{Hirai}, Y., {Wanajo}, S., \& {Saitoh}, T.~R. 2019, arXiv e-prints,
  arXiv:1909.09163

\bibitem[{{Honda} {et~al.}(2007){Honda}, {Aoki}, {Ishimaru}, \&
  {Wanajo}}]{Honda07}
{Honda}, S., {Aoki}, W., {Ishimaru}, Y., \& {Wanajo}, S. 2007, \apj, 666, 1189

\bibitem[{{Honda} {et~al.}(2006){Honda}, {Aoki}, {Ishimaru}, {Wanajo}, \&
  {Ryan}}]{Honda06}
{Honda}, S., {Aoki}, W., {Ishimaru}, Y., {Wanajo}, S., \& {Ryan}, S.~G. 2006,
  \apj, 643, 1180

\bibitem[{{Honda} {et~al.}(2004){Honda}, {Aoki}, Kajino, Ando, Beers, Izumiura,
  Sadakane, \& Takada-Hidai}]{Honda04}
{Honda}, S., {Aoki}, W., Kajino, T., {et~al.} 2004, ApJ, 607, 474

\bibitem[{{Ishimaru} \& {Wanajo}(1999)}]{Ishimaru99}
{Ishimaru}, Y. \& {Wanajo}, S. 1999, \apjl, 511, L33

\bibitem[{{Ishimaru} {et~al.}(2004){Ishimaru}, {Wanajo}, {Aoki}, \&
  {Ryan}}]{Ishimaru04}
{Ishimaru}, Y., {Wanajo}, S., {Aoki}, W., \& {Ryan}, S.~G. 2004, \apj, 600, L47

\bibitem[{{Ishimaru} {et~al.}(2005){Ishimaru}, {Wanajo}, {Aoki}, {Ryan}, \&
  {Prantzos}}]{Ishimaru05}
{Ishimaru}, Y., {Wanajo}, S., {Aoki}, W., {Ryan}, S.~G., \& {Prantzos}, N.
  2005, \nphysa, 758, 603

\bibitem[{{Iwamoto} {et~al.}(1999){Iwamoto}, {Brachwitz}, {Nomoto},
  {Kishimoto}, {Umeda}, {Hix}, \& {Thielemann}}]{Iwamoto99}
{Iwamoto}, K., {Brachwitz}, F., {Nomoto}, K., {et~al.} 1999, \apjs, 125, 439

\bibitem[{{Jablonka} {et~al.}(2015){Jablonka}, {North}, {Mashonkina}, {Hill},
  {Revaz}, {Shetrone}, {Starkenburg}, {Irwin}, {Tolstoy}, {Battaglia}, {Venn},
  {Helmi}, {Primas}, \& {Fran{\c c}ois}}]{Jablonka15}
{Jablonka}, P., {North}, P., {Mashonkina}, L., {et~al.} 2015, \aap, 583, A67

\bibitem[{{Jones} {et~al.}(2016){Jones}, {Ritter}, {Herwig}, {Fryer},
  {Pignatari}, {Bertolli}, \& {Paxton}}]{Jones16}
{Jones}, S., {Ritter}, C., {Herwig}, F., {et~al.} 2016, \mnras, 455, 3848

\bibitem[{{Karakas} \& {Lattanzio}(2014)}]{Karakas14}
{Karakas}, A.~I. \& {Lattanzio}, J.~C. 2014, \pasa, 31, e030

\bibitem[{{Kirby} \& {Cohen}(2012)}]{Kirby12}
{Kirby}, E.~N. \& {Cohen}, J.~G. 2012, \aj, 144, 168

\bibitem[{{Kirby} {et~al.}(2011){Kirby}, {Cohen}, {Smith}, {Majewski}, {Sohn},
  \& {Guhathakurta}}]{Kirby11}
{Kirby}, E.~N., {Cohen}, J.~G., {Smith}, G.~H., {et~al.} 2011, \apj, 727, 79

\bibitem[{{Kobayashi} {et~al.}(2006){Kobayashi}, {Umeda}, {Nomoto}, {Tominaga},
  \& {Ohkubo}}]{Kobayashi06}
{Kobayashi}, C., {Umeda}, H., {Nomoto}, K., {Tominaga}, N., \& {Ohkubo}, T.
  2006, \apj, 653, 1145

\bibitem[{{Koch} {et~al.}(2019){Koch}, {Reichert}, {Hansen}, {Hampel},
  {Stancliffe}, {Karakas}, \& {Arcones}}]{Koch19}
{Koch}, A., {Reichert}, M., {Hansen}, C.~J., {et~al.} 2019, \aap, 622, A159

\bibitem[{{Korobkin} {et~al.}(2012){Korobkin}, {Rosswog}, {Arcones}, \&
  {Winteler}}]{Korobkin12}
{Korobkin}, O., {Rosswog}, S., {Arcones}, A., \& {Winteler}, C. 2012, \mnras,
  426, 1940

\bibitem[{{Langer} {et~al.}(1989){Langer}, {Arcoragi}, \& {Arnould}}]{Langer89}
{Langer}, N., {Arcoragi}, J.~P., \& {Arnould}, M. 1989, \aap, 210, 187

\bibitem[{{Lattimer} {et~al.}(1977){Lattimer}, {Mackie}, {Ravenhall}, \&
  {Schramm}}]{Lattimer77}
{Lattimer}, J.~M., {Mackie}, F., {Ravenhall}, D.~G., \& {Schramm}, D.~N. 1977,
  \apj, 213, 225

\bibitem[{{Limongi} \& {Chieffi}(2018)}]{Limongi18}
{Limongi}, M. \& {Chieffi}, A. 2018, \apjs, 237, 13

\bibitem[{{Lucatello} {et~al.}(2005){Lucatello}, {Tsangarides}, {Beers},
  {Carretta}, {Gratton}, \& {Ryan}}]{Lucatello05}
{Lucatello}, S., {Tsangarides}, S., {Beers}, T.~C., {et~al.} 2005, \apj, 625,
  825

\bibitem[{{Lugaro} {et~al.}(2009){Lugaro}, {Campbell}, \& {de Mink}}]{Lugaro09}
{Lugaro}, M., {Campbell}, S.~W., \& {de Mink}, S.~E. 2009, \pasa, 26, 322

\bibitem[{{Lugaro} {et~al.}(2012){Lugaro}, {Karakas}, {Stancliffe}, \&
  {Rijs}}]{Lugaro12}
{Lugaro}, M., {Karakas}, A.~I., {Stancliffe}, R.~J., \& {Rijs}, C. 2012, \apj,
  747, 2

\bibitem[{{M.~Kovalev} {et~al.}(2018){M.~Kovalev}, {S.~Brinkmann},
  {M.~Bergemann}, \& {MPIA IT-department}}]{NLTE_MPIA}
{M.~Kovalev}, {S.~Brinkmann}, {M.~Bergemann}, \& {MPIA IT-department}. 2018,
  {NLTE MPIA web server, [Online]. Available: {{http://nlte.mpia.de}} Max
  Planck Institute for Astronomy, Heidelberg.}

\bibitem[{{Mashonkina} {et~al.}(2017){Mashonkina}, {Jablonka}, {Sitnova},
  {Pakhomov}, \& {North}}]{Mashonkina17}
{Mashonkina}, L., {Jablonka}, P., {Sitnova}, T., {Pakhomov}, Y., \& {North}, P.
  2017, \aap, 608, A89

\bibitem[{{Mathews} {et~al.}(1992){Mathews}, {Bazan}, \& {Cowan}}]{Mathews92}
{Mathews}, G.~J., {Bazan}, G., \& {Cowan}, J.~J. 1992, \apj, 391, 719

\bibitem[{{Mathews} \& {Cowan}(1990)}]{Mathews90}
{Mathews}, G.~J. \& {Cowan}, J.~J. 1990, \nat, 345, 491

\bibitem[{{McWilliam}(1998)}]{McWilliam98}
{McWilliam}, A. 1998, \aj, 115, 1640

\bibitem[{{Meynet} {et~al.}(2006){Meynet}, {Ekstr{\"o}m}, \&
  {Maeder}}]{MeynetMaeder06}
{Meynet}, G., {Ekstr{\"o}m}, S., \& {Maeder}, A. 2006, \aap, 447, 623

\bibitem[{{Mishenina} {et~al.}(2013){Mishenina}, {Pignatari}, {Korotin},
  {Soubiran}, {Charbonnel}, {Thielemann}, {Gorbaneva}, \&
  {Basak}}]{Mishenina13}
{Mishenina}, T.~V., {Pignatari}, M., {Korotin}, S.~A., {et~al.} 2013, \aap,
  552, A128

\bibitem[{{Montes} {et~al.}(2007){Montes}, {Beers}, {Cowan}, {Elliot},
  {Farouqi}, {Gallino}, {Heil}, {Kratz}, {Pfeiffer}, \& {Pignatari}}]{Montes07}
{Montes}, F., {Beers}, T.~C., {Cowan}, J., {et~al.} 2007, \apj, 671, 1685

\bibitem[{{Nishimura} {et~al.}(2017){Nishimura}, {Sawai}, {Takiwaki}, {Yamada},
  \& {Thielemann}}]{Nishimura17}
{Nishimura}, N., {Sawai}, H., {Takiwaki}, T., {Yamada}, S., \& {Thielemann},
  F.~K. 2017, \apj, 836, L21

\bibitem[{{Nishimura} {et~al.}(2015){Nishimura}, {Takiwaki}, \&
  {Thielemann}}]{Nishimura15}
{Nishimura}, N., {Takiwaki}, T., \& {Thielemann}, F.-K. 2015, \apj, 810, 109

\bibitem[{{Norris} \& {Yong}(2019)}]{Norris19}
{Norris}, J.~E. \& {Yong}, D. 2019, \apj, 879, 37

\bibitem[{{Norris} {et~al.}(2013){Norris}, {Yong}, {Bessell}, {Christlieb},
  {Asplund}, {Gilmore}, {Wyse}, {Beers}, {Barklem}, {Frebel}, \&
  {Ryan}}]{Norris13}
{Norris}, J.~E., {Yong}, D., {Bessell}, M.~S., {et~al.} 2013, \apj, 762, 28

\bibitem[{{North} {et~al.}(2012){North}, {Cescutti}, {Jablonka}, {Hill},
  {Shetrone}, {Letarte}, {Lemasle}, {Venn}, {Battaglia}, {Tolstoy}, {Irwin},
  {Primas}, \& {Fran{\c c}ois}}]{North12}
{North}, P., {Cescutti}, G., {Jablonka}, P., {et~al.} 2012, \aap, 541, A45

\bibitem[{{Pignatari} {et~al.}(2008){Pignatari}, {Gallino}, {Meynet},
  {Hirschi}, {Herwig}, \& {Wiescher}}]{Pignatari08}
{Pignatari}, M., {Gallino}, R., {Meynet}, G., {et~al.} 2008, \apj, 687, L95

\bibitem[{{Placco} {et~al.}(2014){Placco}, {Frebel}, {Beers}, \&
  {Stancliffe}}]{Placco14}
{Placco}, V.~M., {Frebel}, A., {Beers}, T.~C., \& {Stancliffe}, R.~J. 2014,
  \apj, 797, 21

\bibitem[{{Prantzos} {et~al.}(1990){Prantzos}, {Hashimoto}, \&
  {Nomoto}}]{Prantzos90}
{Prantzos}, N., {Hashimoto}, M., \& {Nomoto}, K. 1990, \aap, 234, 211

\bibitem[{{Qian} \& {Wasserburg}(2003)}]{Qian03}
{Qian}, Y.~Z. \& {Wasserburg}, G.~J. 2003, \apj, 588, 1099

\bibitem[{{Raiteri} {et~al.}(1991){Raiteri}, {Busso}, {Gallino}, \&
  {Picchio}}]{Raiteri91}
{Raiteri}, C.~M., {Busso}, M., {Gallino}, R., \& {Picchio}, G. 1991, \apj, 371,
  665

\bibitem[{{Reddy} {et~al.}(2006){Reddy}, {Lambert}, \& {Allende
  Prieto}}]{Reddy06}
{Reddy}, B.~E., {Lambert}, D.~L., \& {Allende Prieto}, C. 2006, \mnras, 367,
  1329

\bibitem[{{Reddy} {et~al.}(2003){Reddy}, {Tomkin}, {Lambert}, \& {Allende
  Prieto}}]{Reddy03}
{Reddy}, B.~E., {Tomkin}, J., {Lambert}, D.~L., \& {Allende Prieto}, C. 2003,
  \mnras, 340, 304

\bibitem[{{Reggiani} {et~al.}(2019){Reggiani}, {Amarsi}, {Lind}, {Barklem},
  {Zatsarinny}, {Bartschat}, {Fursa}, {Bray}, {Spina}, \&
  {Mel{\'e}ndez}}]{Reggiani19}
{Reggiani}, H., {Amarsi}, A.~M., {Lind}, K., {et~al.} 2019, \aap, 627, A177

\bibitem[{{Roederer} {et~al.}(2010){Roederer}, {Cowan}, {Karakas}, {Kratz},
  {Lugaro}, {Simmerer}, {Farouqi}, \& {Sneden}}]{Roederer10}
{Roederer}, I.~U., {Cowan}, J.~J., {Karakas}, A.~I., {et~al.} 2010, \apj, 724,
  975

\bibitem[{{Roederer} {et~al.}(2016){Roederer}, {Karakas}, {Pignatari}, \&
  {Herwig}}]{Roederer16}
{Roederer}, I.~U., {Karakas}, A.~I., {Pignatari}, M., \& {Herwig}, F. 2016,
  \apj, 821, 37

\bibitem[{{Roederer} {et~al.}(2014){Roederer}, {Preston}, {Thompson},
  {Shectman}, {Sneden}, {Burley}, \& {Kelson}}]{Roederer14}
{Roederer}, I.~U., {Preston}, G.~W., {Thompson}, I.~B., {et~al.} 2014, \aj,
  147, 136

\bibitem[{{Rosswog} {et~al.}(1999){Rosswog}, {Liebend{\"o}rfer}, {Thielemann},
  {Davies}, {Benz}, \& {Piran}}]{Rosswog99}
{Rosswog}, S., {Liebend{\"o}rfer}, M., {Thielemann}, F.~K., {et~al.} 1999,
  \aap, 341, 499

\bibitem[{{Rosswog} {et~al.}(2013){Rosswog}, {Piran}, \& {Nakar}}]{Rosswog13}
{Rosswog}, S., {Piran}, T., \& {Nakar}, E. 2013, \mnras, 430, 2585

\bibitem[{{Salvadori} {et~al.}(2019){Salvadori}, {Bonifacio}, {Caffau},
  {Korotin}, {Andreevsky}, {Spite}, \& {Sk{\'u}lad{\'o}ttir}}]{Salvadori19}
{Salvadori}, S., {Bonifacio}, P., {Caffau}, E., {et~al.} 2019, \mnras, 487,
  4261

\bibitem[{{Salvadori} {et~al.}(2015){Salvadori}, {Sk{\'u}lad{\'o}ttir}, \&
  {Tolstoy}}]{Salvadori15}
{Salvadori}, S., {Sk{\'u}lad{\'o}ttir}, {\'A}., \& {Tolstoy}, E. 2015, \mnras,
  454, 1320

\bibitem[{{Savino} {et~al.}(2018){Savino}, {de Boer}, {Salaris}, \&
  {Tolstoy}}]{Savino18}
{Savino}, A., {de Boer}, T.~J.~L., {Salaris}, M., \& {Tolstoy}, E. 2018,
  \mnras, 480, 1587

\bibitem[{{Shetrone} {et~al.}(2003){Shetrone}, {Venn}, {Tolstoy}, {Primas},
  {Hill}, \& {Kaufer}}]{Shetrone03}
{Shetrone}, M., {Venn}, K.~A., {Tolstoy}, E., {et~al.} 2003, \aj, 125, 684

\bibitem[{{Siegel}(2019)}]{Siegel19b}
{Siegel}, D.~M. 2019, arXiv e-prints, arXiv:1901.09044

\bibitem[{{Siegel} {et~al.}(2019){Siegel}, {Barnes}, \& {Metzger}}]{Siegel19a}
{Siegel}, D.~M., {Barnes}, J., \& {Metzger}, B.~D. 2019, \nat, 569, 241

\bibitem[{{Simmerer} {et~al.}(2004){Simmerer}, {Sneden}, {Cowan}, {Collier},
  {Woolf}, \& {Lawler}}]{Simmerer04}
{Simmerer}, J., {Sneden}, C., {Cowan}, J.~J., {et~al.} 2004, \apj, 617, 1091

\bibitem[{{Sk{\'u}lad{\'o}ttir}
  {et~al.}(2015{\natexlab{a}}){Sk{\'u}lad{\'o}ttir}, {Andrievsky}, {Tolstoy},
  {Hill}, {Salvadori}, {Korotin}, \& {Pettini}}]{Skuladottir15b}
{Sk{\'u}lad{\'o}ttir}, {\'A}., {Andrievsky}, S.~M., {Tolstoy}, E., {et~al.}
  2015{\natexlab{a}}, \aap, 580, A129

\bibitem[{{Sk{\'u}lad{\'o}ttir} {et~al.}(2019){Sk{\'u}lad{\'o}ttir}, {Hansen},
  {Salvadori}, \& {Choplin}}]{Skuladottir19}
{Sk{\'u}lad{\'o}ttir}, {\'A}., {Hansen}, C.~J., {Salvadori}, S., \& {Choplin},
  A. 2019, \aap, 631, A171

\bibitem[{{Sk{\'u}lad{\'o}ttir} {et~al.}(2018){Sk{\'u}lad{\'o}ttir},
  {Salvadori}, {Pettini}, {Tolstoy}, \& {Hill}}]{Skuladottir18}
{Sk{\'u}lad{\'o}ttir}, {\'A}., {Salvadori}, S., {Pettini}, M., {Tolstoy}, E.,
  \& {Hill}, V. 2018, \aap, 615, A137

\bibitem[{{Sk{\'u}lad{\'o}ttir} {et~al.}(2017){Sk{\'u}lad{\'o}ttir}, {Tolstoy},
  {Salvadori}, {Hill}, \& {Pettini}}]{Skuladottir17}
{Sk{\'u}lad{\'o}ttir}, {\'A}., {Tolstoy}, E., {Salvadori}, S., {Hill}, V., \&
  {Pettini}, M. 2017, \aap, 606, A71

\bibitem[{{Sk{\'u}lad{\'o}ttir}
  {et~al.}(2015{\natexlab{b}}){Sk{\'u}lad{\'o}ttir}, {Tolstoy}, {Salvadori},
  {Hill}, {Pettini}, {Shetrone}, \& {Starkenburg}}]{Skuladottir15a}
{Sk{\'u}lad{\'o}ttir}, {\'A}., {Tolstoy}, E., {Salvadori}, S., {et~al.}
  2015{\natexlab{b}}, \aap, 574, A129

\bibitem[{{Sneden} {et~al.}(2008){Sneden}, {Cowan}, \& {Gallino}}]{Sneden08}
{Sneden}, C., {Cowan}, J.~J., \& {Gallino}, R. 2008, \araa, 46, 241

\bibitem[{{Sneden} {et~al.}(2003){Sneden}, {Cowan}, {Lawler}, {Ivans},
  {Burles}, {Beers}, {Primas}, {Hill}, {Truran}, {Fuller}, {Pfeiffer}, \&
  {Kratz}}]{Sneden03}
{Sneden}, C., {Cowan}, J.~J., {Lawler}, J.~E., {et~al.} 2003, \apj, 591, 936

\bibitem[{{Spite} {et~al.}(2005){Spite}, {Cayrel}, {Plez}, {Hill}, {Spite},
  {Depagne}, {Fran{\c{c}}ois}, {Bonifacio}, {Barbuy}, {Beers}, {Andersen},
  {Molaro}, {Nordstr{\"o}m}, \& {Primas}}]{Spite05}
{Spite}, M., {Cayrel}, R., {Plez}, B., {et~al.} 2005, \aap, 430, 655

\bibitem[{{Spite} {et~al.}(2018){Spite}, {Spite}, {Fran{\c{c}}ois},
  {Bonifacio}, {Caffau}, \& {Salvadori}}]{Spite18}
{Spite}, M., {Spite}, F., {Fran{\c{c}}ois}, P., {et~al.} 2018, \aap, 617, A56

\bibitem[{{Starkenburg} {et~al.}(2014){Starkenburg}, {Shetrone}, {McConnachie},
  \& {Venn}}]{Starkenburg14}
{Starkenburg}, E., {Shetrone}, M.~D., {McConnachie}, A.~W., \& {Venn}, K.~A.
  2014, \mnras, 441, 1217

\bibitem[{{Suda} {et~al.}(2008){Suda}, {Katsuta}, {Yamada}, {Suwa}, {Ishizuka},
  {Komiya}, {Sorai}, {Aikawa}, \& {Fujimoto}}]{Suda08}
{Suda}, T., {Katsuta}, Y., {Yamada}, S., {et~al.} 2008, \pasj, 60, 1159

\bibitem[{{Susmitha} {et~al.}(2017){Susmitha}, {Koch}, \&
  {Sivarani}}]{Susmitha17}
{Susmitha}, A., {Koch}, A., \& {Sivarani}, T. 2017, \aap, 606, A112

\bibitem[{{Tolstoy} {et~al.}(2009){Tolstoy}, {Hill}, \& {Tosi}}]{Tolstoy09}
{Tolstoy}, E., {Hill}, V., \& {Tosi}, M. 2009, \araa, 47, 371

\bibitem[{{Travaglio} {et~al.}(2004){Travaglio}, {Gallino}, {Arnone}, {Cowan},
  {Jordan}, \& {Sneden}}]{Travaglio04}
{Travaglio}, C., {Gallino}, R., {Arnone}, E., {et~al.} 2004, \apj, 601, 864

\bibitem[{{Tsujimoto} {et~al.}(1995){Tsujimoto}, {Nomoto}, {Yoshii},
  {Hashimoto}, {Yanagida}, \& {Thielemann}}]{Tsujimoto95}
{Tsujimoto}, T., {Nomoto}, K., {Yoshii}, Y., {et~al.} 1995, \mnras, 277, 945

\bibitem[{{Umeda} \& {Nomoto}(2003)}]{UmedaNomoto03}
{Umeda}, H. \& {Nomoto}, K. 2003, \nat, 422, 871

\bibitem[{{Venn} {et~al.}(2004){Venn}, {Irwin}, {Shetrone}, {Tout}, {Hill}, \&
  {Tolstoy}}]{Venn04}
{Venn}, K.~A., {Irwin}, M., {Shetrone}, M.~D., {et~al.} 2004, \aj, 128, 1177

\bibitem[{{Wanajo} {et~al.}(2011){Wanajo}, {Janka}, \& {M{\"u}ller}}]{Wanajo11}
{Wanajo}, S., {Janka}, H.-T., \& {M{\"u}ller}, B. 2011, \apj, 726, L15

\bibitem[{{Wanajo} {et~al.}(2001){Wanajo}, {Kajino}, {Mathews}, \&
  {Otsuki}}]{Wanajo01}
{Wanajo}, S., {Kajino}, T., {Mathews}, G.~J., \& {Otsuki}, K. 2001, \apj, 554,
  578

\bibitem[{{Wanajo} {et~al.}(2009){Wanajo}, {Nomoto}, {Janka}, {Kitaura}, \&
  {M{\"u}ller}}]{Wanajo09}
{Wanajo}, S., {Nomoto}, K., {Janka}, H.~T., {Kitaura}, F.~S., \& {M{\"u}ller},
  B. 2009, \apj, 695, 208

\bibitem[{{Wanajo} {et~al.}(2003){Wanajo}, {Tamamura}, {Itoh}, {Nomoto},
  {Ishimaru}, {Beers}, \& {Nozawa}}]{Wanajo03}
{Wanajo}, S., {Tamamura}, M., {Itoh}, N., {et~al.} 2003, \apj, 593, 968

\bibitem[{{Watson} {et~al.}(2019){Watson}, {Hansen}, {Selsing}, {Koch},
  {Malesani}, {Andersen}, {Fynbo}, {Arcones}, {Bauswein}, {Covino}, {Grado},
  {Heintz}, {Hunt}, {Kouveliotou}, {Leloudas}, {Levan}, {Mazzali}, \&
  {Pian}}]{Watson19}
{Watson}, D., {Hansen}, C.~J., {Selsing}, J., {et~al.} 2019, \nat, 574, 497

\bibitem[{{Weisz} {et~al.}(2014){Weisz}, {Dolphin}, {Skillman}, {Holtzman},
  {Gilbert}, {Dalcanton}, \& {Williams}}]{Weisz14}
{Weisz}, D.~R., {Dolphin}, A.~E., {Skillman}, E.~D., {et~al.} 2014, \apj, 789,
  147

\bibitem[{{Wheeler} {et~al.}(1998){Wheeler}, {Cowan}, \&
  {Hillebrandt}}]{Wheeler98}
{Wheeler}, J.~C., {Cowan}, J.~J., \& {Hillebrandt}, W. 1998, \apjl, 493, L101

\bibitem[{{Winteler} {et~al.}(2012){Winteler}, {K{\"a}ppeli}, {Perego},
  {Arcones}, {Vasset}, {Nishimura}, {Liebend{\"o}rfer}, \&
  {Thielemann}}]{Winteler12}
{Winteler}, C., {K{\"a}ppeli}, R., {Perego}, A., {et~al.} 2012, \apjl, 750, L22

\bibitem[{{Yong} {et~al.}(2017){Yong}, {Norris}, {Da Costa}, {Stanford},
  {Karakas}, {Shingles}, {Hirschi}, \& {Pignatari}}]{Yong17}
{Yong}, D., {Norris}, J.~E., {Da Costa}, G.~S., {et~al.} 2017, \apj, 837, 176

\end{thebibliography}

\end{document}